\documentclass[aps,pre, amsmath,amssymb,preprint,groupedaddress]{revtex4-2}

\usepackage{booktabs}
\usepackage{subcaption}
\usepackage{hyperref}
\usepackage{graphicx}
\usepackage{dcolumn}
\usepackage{bm}
\usepackage[mathlines]{lineno}

\usepackage{xcolor}
\newcommand{\rev}[1]{{\color{black}#1}}

\newcommand{\Yb}{\bar{Y}}

\begin{document}

\title{Linking Dynamical Time Scales and Experimental Design to Parameter Identifiability from Time-Series Observations}

\author{Sara Shamekh}
\email[]{ss18284@nyu.edu}

\affiliation{Courant Institute School of Mathematics, Computing, and Data Science, New York University, New York, NY, USA}
\author{Viviana Acquaviva}

\affiliation{CUNY New York City College of Technology, Brooklyn, New York, NY, USA}
\affiliation{Lamont-Doherty Earth Observatory, Columbia University, Palisades, NY, USA}

\begin{abstract}
\rev{Parameter estimation in chaotic models often relies on prescribed summary statistics, whose ability to constrain parameters may be difficult to establish in advance. Different parameters influence different features and time scales of the dynamics, so the record length and sampling interval needed to constrain them may differ.  We choose to learn a representation directly from the trajectory and use it as our summary statistic.  The recurrent encoder that produces this representation is trained jointly with a conditional normalizing flow for amortized posterior estimation. We use the Lorenz-63 and two-scale Lorenz-96 systems to investigate how inference depends on the observed variables, record length and sampling interval, and compare our results  with the state-of-the-art Calibrate--Emulate--Sample method.  In Lorenz-96, learned summaries of a 2 model-time-unit (MTU) record achieve estimation  skill comparable to prescribed second-moment statistics from substantially longer records for several parameters.  Fast-process parameters remain weakly constrained by slow-variable records across the tested 2 to 100 MTU windows, while adding fast-field averages improves their recovery from 2 MTU. Coarser sampling weakens inference from fast-field observations, but longer coverage can partly compensate.  
Once trained for a given observation setting, the estimator we develop returns posteriors for compatible records without further model integration. These proof-of-concept results illustrate how data-driven learning of summaries  can be used to inform the design of observations in dynamical systems.}
\end{abstract}

\maketitle
\section{Introduction}
\label{sec:Intro}

\rev{In parameter estimation, a physical model links unknown parameters to observable quantities, and the inverse problem is to infer those parameters from observations.}
In dynamical systems, such inference has traditionally been addressed either deterministically, through optimization-based calibration \cite{tarantola2005inverse}, or probabilistically, within a Bayesian framework using sampling methods, particle filters, or Kalman-based approaches \cite{kaipio2005statistical,welch1995introduction}.
Inference in dynamical systems is inherently a time-series problem, since
observations from instruments, reanalyses, or numerical simulators are collected as trajectories that encode not only mean behavior, but also variability, persistence, intermittency, and transitions across multiple time scales.
\rev{Observed variables, record length and sampling interval shape the parameter information available for inference.}
 Many scientific simulators evolve state variables
\(x(t)\in\mathbb{R}^d\) according to nonlinear dynamics governed by tunable
parameters \(\theta\in\Theta\subset\mathbb{R}^p\),
\[
\dot{x}(t)=F\big(x(t);\theta\big),
\]
with observations \(y(t)\in\mathbb{R}^m\) corresponding either to the full state
or to a partial, noisy transformation
\[
y(t)=\mathcal{H}\!\big(x(t)\big)+\varepsilon(t),
\]
where $\mathcal{H}$ is the observation operator and $\varepsilon(t)$ represents measurement noise. The calibration problem is to infer \(\theta\) from a finite observed trajectory
\(y_{\mathrm{obs}}\) so that simulations reproduce the observed behavior.

A natural approach is to compare simulated and observed trajectories directly.
In chaotic systems, however, small differences in parameters or initial
conditions lead to rapid dephasing, so trajectories that are dynamically
consistent diverge pointwise in time \cite{lorenz63,abarbanel2013predicting,jarvinen2012ensemble,schneider2017earth}.
Pointwise misfit  penalizes phase error rather than differences in the underlying dynamics, making direct trajectory matching an unreliable basis for parameter inference.
This limitation is intrinsic to chaos and motivates inference strategies that do not rely on exact temporal alignment.

To bypass trajectory misalignment and reduce computational cost, simulator outputs and observations are often compressed into hand-crafted summary statistics
\(S(y)\), such as time-averaged profiles, variances, covariances, extremes, or other low-order aggregates, and parameters are inferred by matching these
summaries using likelihoods or distances (e.g., synthetic
likelihood \cite{ong2018likelihood,price2018bayesian}, approximate Bayesian computation (ABC) \cite{Beaumont2002_PNAS,sisson2018handbook}).
Emulators are often introduced to further accelerate inference, either by
approximating the forward map \(\theta\mapsto \hat y\) or directly the summary map \(\theta\mapsto \widehat{S(y)}\), followed by MCMC or optimization on the
surrogate \cite{hess2024emulator,yang2025emulator,keetz2025emulator,Cleary2021_CES_JCP}.
\rev{Physically motivated summaries can provide strong parameter constraints and form the basis of established calibration methods, even when their sufficiency is unknown. We use such statistics as benchmarks for learned representations.}

From a statistical perspective, preserving all posterior information requires that the chosen summaries \(S(y)\) be sufficient for the parameter vector \(\theta\), although calibration can also use information-losing summaries.
While sufficient statistics may be known explicitly for simple models, there is no general procedure for identifying them in complex nonlinear systems.
In practice, summaries are  chosen for convenience or interpretability and are often limited to time-averaged profiles or low-order moments.
As a result, distinct parameter values may reproduce the same summaries while differing substantially in their temporal organization, leading to practical
non-identifiability and strongly correlated, multimodal, or weakly constrained posteriors\rev{, an issue} that is pervasive in climate, fluid, and ecological modeling, where parameters act across multiple dynamical time scales \cite{Dunbar_2021,Lguensat_2023,Herrera2022,Lederman2022}. \rev{A trajectory contains at least as much parameter information as any deterministic summary of it. Whether a trained estimator extracts that information is an empirical question, which we assess through matched-record comparisons.}

\rev{We ask whether informative summaries can be learned directly from time series, reducing reliance on choosing statistics in advance. We evaluate learned representations against established, physically motivated summaries and examine how their gains depend on observed variables, window length $T$ and sampling interval $\Delta t$. These choices determine which processes are recorded and how well their variability is resolved.}

\rev{We combine representation learning with amortized posterior estimation.} Given a trajectory \(y_{0{:}T}\), we construct a learned representation
\[
h = f_\psi(y_{0{:}T}),
\]
which serves as a data-adaptive compression of the trajectory.
Unlike time-averaged summaries, \(h\) is free to encode time-lagged dependencies, phase relationships, and multiscale structure, without requiring explicit
alignment in time.
\rev{Training through the inference objective encourages $h$ to retain features informative about $\theta$.}

Following Ref.~\cite{Radev2023_BayesFlow}, we adopt a posterior-centric, likelihood-free inference framework in which the goal is to approximate \(p(\theta \mid y_{0{:}T})\) directly.
A conditional normalizing flow models the posterior distribution
\(q_\phi(\theta \mid h)\) given the learned representation \(h\), yielding a
flexible, invertible density capable of representing  correlations and
multimodality.
The encoder \(f_\psi\) and the normalizing flow are trained jointly on simulated pairs
\((\theta,y_{0{:}T})\).
Once training is complete, posterior inference for new trajectories requires only a single forward pass through the trained model.
This amortized setting allows existing simulation datasets to be reused efficiently, without additional forward-model evaluations \cite{Radev2023_BayesFlow,zammitmangion2024amortized}.
In particular, one natural application would be the  large perturbed-parameter ensembles that are routinely generated in climate model calibration at significant computational cost \cite{eidhammer2024ppe,Elsaesser25PPE,sexton2021perturbed}.
\rev{A trained estimator can serve many observational records compatible with its model, prior and observation setting. Such reuse could support repeated climate calibration across windows, sites or datasets.}

\rev{To compare observation designs}, we train separate amortized inference models for each combination of \rev{observed variables,} \(T\) and \(\Delta t\). Each model is trained on trajectories generated at a fixed temporal window and sampling resolution, and learns a posterior map tailored to that observational design.
This controlled setup allows us to attribute changes in posterior uncertainty, correlations, and multimodality directly to the \rev{observed variables and the} temporal coverage and resolution of the data, rather than to model mismatch\rev{. This attribution remains subject to estimation error. We thus train three seeds per configuration and compare trajectory-conditioned flows with flows conditioned on functions of the same record.} By comparing posteriors across \rev{designs}, we obtain a parameter-wise characterization of identifiability and link observed trends to the dynamical time scales through which parameters shape the system's evolution.

Several recent studies have exploited time-series structure directly for
parameter estimation in dynamical systems by conditioning inference on learned
temporal representations rather than hand-crafted summaries.
For example, \citet{rudi2022parameter} use neural encodings of trajectories to
infer parameters in nonlinear dynamical systems and show that the encoding learned from the trajectories can substantially improve identifiability.
Related simulation-based inference approaches similarly rely on recurrent or
convolutional encoders to map time series to posterior distributions or
likelihood surrogates in chaotic and stochastic systems
\cite{Papamakarios2019_SNLE,Lueckmann2021_BenchmarkSBI,Radev2023_BayesFlow}.
These works demonstrate the value of learning from full trajectories, but
typically assume a fixed temporal sampling of the data. \rev{We build on these approaches by jointly examining representation choice and observation design, and by comparing with Calibrate--Emulate--Sample, an established framework for calibration from time-averaged statistics \cite{Cleary2021_CES_JCP,Dunbar_2021}. Section~\ref{sec:ces} evaluates both posterior quality and computational cost.}

\section{Problem setup}

\subsection{Dynamical system and observation model}

We consider parameter estimation for the nonlinear dynamical systems introduced
in Section~\ref{sec:Intro}, with state evolution governed by
\(\dot{x}(t)=F(x(t);\theta)\).
Here, \(x(t)\) denotes the system state and \(\theta\) the vector of unknown model
parameters, whose components may act through distinct intrinsic time scales.

Observations consist of a finite time series obtained from the state via the
observation operator \(\mathcal{H}\), possibly corrupted by noise,
\(y(t)=\mathcal{H}(x(t))+\varepsilon(t)\).
We allow for partial and noisy observations, and do not assume that
\(\mathcal{H}\) is invertible; consequently, different states or parameter values
may be consistent with the same observed trajectory.

 \rev{Baseline experiments use noise-free observations, $\varepsilon = 0$. Noise sensitivity is reported in Appendix~\ref{sec:app_noise}. Lorenz-63 is fully observed. For two-scale Lorenz-96, we observe slow variables $X$, fast variables $Y$, or $X$ together with fast-field site averages $\Yb_k = J^{-1}\sum_j Y_{j,k}$. The last setting represents archiving resolved fields alongside an aggregate diagnostic of a subgrid process.}

Over a finite observation window of duration \(T\) and sampling interval
\(\Delta t\), the available data consist of the discrete trajectory
\begin{equation}
\label{eq:traj}
y_{0{:}T} = \{ y(t_0), y(t_0+\Delta t), \ldots, y(t_0+N\Delta t) \},
\qquad N\Delta t = T,
\end{equation}
which we treat as the input to the inference problem.
Our goal is to characterize the posterior distribution
\begin{equation}
\label{eq:posterior}
p(\theta \mid y_{0{:}T}),
\end{equation}
which may be non-Gaussian, correlated, or multimodal, and whose concentration
depends on the temporal coverage \(T\) and the sampling resolution \(\Delta t\).

Although the posterior of interest is formally defined as
\(p(\theta \mid y_{0{:}T})\), inference is performed by conditioning on a learned
representation \(h = f_\psi(y_{0{:}T})\).
This substitution can be justified through the lens of sufficient statistics,
since conditioning on a sufficient statistic preserves the posterior
distribution.
By jointly training the encoder and conditional density model to maximize
\(\log q_\phi(\theta \mid h)\), the learned representation is encouraged
to retain those aspects of the trajectory that are most informative about
\(\theta\).
If a sufficient statistic exists within the expressive capacity of the
encoder--flow family, training may recover it, but this is not guaranteed.
Training seeks to minimize the expected negative log posterior density
\(\mathbb{E}[-\log q_\phi(\theta \mid h)]\), without guaranteeing global optimality.
Sufficiency does not imply identifiability, since a sufficient representation
can preserve a broad posterior for weakly constrained parameters.

\subsection{Temporal window length and sampling resolution}
\label{sec:design}

\rev{Learned summaries depend on the observations from which they are constructed. We vary window length $T$ and sampling interval $\Delta t$ separately to distinguish temporal coverage from resolution.

\paragraph{Window length.}
Longer records can sample more of the attractor, improve estimates of time averages and variability, and capture infrequent transitions. These changes may strengthen constraints on parameters that affect long-term behavior, even with relatively coarse sampling.

\paragraph{Temporal resolution.}
Fine sampling can reveal parameter-sensitive rates and short-time dependencies. Coarser sampling may weaken these constraints without removing all parameter information, and longer coverage may compensate through other features of the dynamics. Observation noise can obscure short-time changes, particularly when consecutive samples differ little relative to the noise amplitude.

\paragraph{Units.}
In the observation equations, $T$ and $\Delta t$ denote physical duration and sampling interval. Tables report windows in model time units (MTU) and also use $T$ and $\Delta t$ for counts of stored steps of $0.01$ MTU. Thus, table entries $T = 200$ and $\Delta t = 20$ denote a $2$ MTU window sampled every $0.2$ MTU. Integration steps are finer, $0.002$ MTU for Lorenz-63 and $0.0005$ MTU for Lorenz-96. Even the finest observed records therefore retain one sample every five integration steps for Lorenz-63 and every twenty steps for Lorenz-96. The estimator receives sampled states without supplied derivatives or numerical-step increments. 
}
\subsection{Amortized posterior inference with conditional normalizing flows}

We approximate the posterior \(p(\theta \mid y_{0{:}T}^{(\Delta t)})\) using an
amortized simulation-based inference framework \rev{based} on conditional generative density estimation, specifically \rev{a} conditional normalizing flow \cite{rezende2015variational,kobyzev2020normalizing,papamakarios2017masked}.

We train a neural conditional density estimator that learns to map trajectories directly to the full posterior distribution over parameters. This eliminates the need for a separate optimization or sampling procedure for each observed trajectory.

Inference proceeds by mapping the observed trajectory
\(y_{0{:}T}^{(\Delta t)}\) to a learned representation
\begin{equation}
\label{eq:encoder}
h = f_\psi\!\left(y_{0{:}T}^{(\Delta t)}\right),
\end{equation}
which serves as a data-adaptive summary of the time series.
The encoder \(f_\psi\) is implemented here as a long short-term memory (LSTM) network,
chosen to capture temporal ordering and long-range dependencies in the trajectory; however, the inference framework itself does
not rely on this specific architecture.

Conditioned on \(h\), the posterior over parameters is modeled using a
conditional normalizing flow,
\begin{equation}
\label{eq:flow}
\theta = g_\phi(z; h), \qquad z \sim p_0(z),
\end{equation}
where \(p_0\) is a simple base distribution and
\(g_\phi(\cdot;h)\) is an invertible, differentiable transformation.
\rev{For fixed $h$, $g_\phi$ is an invertible map from $z$ to $\theta$. Its induced density follows from the change-of-variables formula,
\begin{equation}
\label{eq:cov}
q_\phi(\theta \mid h)
= p_0\!\left(g_\phi^{-1}(\theta;h)\right)\,
\left|\det \frac{\partial g_\phi^{-1}(\theta;h)}{\partial \theta}\right| ,
\qquad p_0 = \mathcal{N}(0,I).
\end{equation}
The Jacobian accounts for local volume changes, ensuring a non-negative, normalized conditional density. Applying $g_\phi$ to a base draw generates a posterior sample, while its inverse and Jacobian evaluate the density at a given $\theta$. We use four masked autoregressive layers \cite{papamakarios2017masked}. Each applies coordinate-wise affine transformations with scales and shifts conditioned on preceding coordinates and $h$, with fixed permutations between layers. Invertibility and triangular Jacobians make density evaluation tractable. This construction can represent skewed, correlated and multimodal posteriors.}

The encoder and flow parameters \((\psi,\phi)\) are trained jointly on simulated
pairs \((\theta, y_{0{:}T}^{(\Delta t)})\) by maximizing the conditional log
likelihood
\begin{equation}
\label{eq:loss}
\mathcal{L}(\psi,\phi)
=
\mathbb{E}_{p(\theta,y)}\!
\left[
  \log q_\phi\bigl(\theta \mid f_\psi(y_{0{:}T}^{(\Delta t)})\bigr)
\right].
\end{equation}
\rev{Equation~\eqref{eq:loss} rewards density assigned to the parameters that generated each trajectory. Normalization prevents increasing this density everywhere. For a fixed encoder, $h = f_\psi(y)$, the negative objective is
\[
H(\theta \mid h)
+
\mathbb{E}_{p(h)}
\left[
D_{\mathrm{KL}}\!\left(
p(\theta \mid h)\,\|\,q_\phi(\theta \mid h)
\right)
\right].
\]
Optimizing $\phi$ reduces the discrepancy from $p(\theta\mid h)$. Jointly optimizing $\psi$ also favors informative representations. In the ideal limit where the flow matches this conditional distribution, maximizing the objective minimizes $H(\theta\mid h)$ and maximizes information retained about $\theta$.

No separate likelihood or error covariance is prescribed for $h$. Conditional variability is learned from simulated parameter--trajectory pairs, so the inferred uncertainty depends on variability represented in those simulations. Training targets parameter inference rather than trajectory reconstruction, encouraging retention of features useful for constraining $\theta$.}

After training, the posterior for a new trajectory is one forward pass through the encoder and flow, so no further forward-model evaluations are needed at inference time.

\rev{\paragraph{Encoder.}
A single-layer LSTM encodes each trajectory. Its temporal mean, temporal maximum and final hidden state are concatenated to form $h$, providing access to pooled and sequential features.

\paragraph{Conditioning.}
We train the same flow architecture on five types of inputs. These are raw trajectories, summaries based on first and second moments, first-moment summaries alone, block-reduced trajectories, and hybrid inputs combining trajectories with their means, standard deviations and correlations. First-moment summaries are the time means of each component of the observation vector. For block reduction, we divide each trajectory into twenty-five blocks and retain each component's mean and standard deviation within each block. Flows conditioned only on prescribed summaries use no encoder. The moment-based benchmarks are established, physically motivated statistics defined for each system in Section~\ref{sec:results}, where we distinguish raw moments from summaries expressed as means, standard deviations and correlations.

\paragraph{Training.}
We split the $1024$ parameter values into disjoint training, validation and test sets in approximate proportions of $70\%$, $15\%$ and $15\%$. All $50$ realizations for a given parameter value stay in the same set. The $718$ training values  yield $35{,}900$ training trajectories. Across configurations, we use windows that start at the beginning of the same stored trajectories.

The LSTM has $80$ units, and each of the four masked autoregressive layers uses a conditioner with $64$ units. We train with Adam using a learning rate of $0.009$, a weight decay of $5\times10^{-5}$ and a batch size of $512$. Training stops if the validation loss does not improve for $13$ consecutive epochs, with a maximum of $300$ epochs. We train each configuration with three random seeds and draw $2000$ posterior samples per test trajectory. We report sensitivity to the training seed in the Results and Appendix~\ref{sec:app_seeds}.

\subsection{Scoring}
\label{sec:scoring}
For each parameter, we evaluate posterior means against the true parameter values across the held-out trajectories. Tables in Section~\ref{sec:results} report the coefficient of determination $R^2$. A negative score means that the posterior means give a larger squared error than predicting the mean of the true values in the test set. We compute posterior means using only draws within the prior bounds. The discarded draws account for about $5\%$ of the samples for Lorenz-96 and a much smaller fraction for Lorenz-63 (Table~\ref{tab:support}).

A deterministic summary, subsample, shorter prefix or subset of observed variables cannot contain more parameter information than the record from which it was derived. A trained estimator can nevertheless perform better on a reduced input. If the estimator consistently does so when both inputs come from the same record, the comparison points to a limitation of the estimator given the full trajectory. We examine available matched-record comparisons in Appendix~\ref{sec:app_check}.}

\subsection{Test cases}
\subsubsection{Lorenz-63 experiment}

\rev{Our first test case is the Lorenz-63 system \cite{lorenz63}, in which the full state is observed and every parameter acts directly on an observed variable.}
The dynamics are governed by
\begin{align}
\dot{x} &= \sigma (y - x), \nonumber\\
\dot{y} &= x(\rho - z) - y, \label{eq:l63}\\
\dot{z} &= xy - \beta z,\nonumber
\end{align}
with state \((x(t), y(t), z(t)) \in \mathbb{R}^3\) and parameter vector
\(\theta = (\rho, \sigma, \beta)\).

\rev{
Parameters are drawn from a Gaussian prior centered on the canonical values $(\rho,\sigma,\beta) = (28, 10, 8/3)$ with standard deviations $(3, 1.5, 0.35)$, truncated to $\rho \in [26,36]$, $\sigma \in [8,14]$ and $\beta \in [2,3]$. Truncation is by rejection, so no mass is piled on the boundary. Every one of the $1024$ prior samples was checked numerically and gives a chaotic trajectory. For each parameter value, $50$ trajectories are integrated from distinct initial conditions to capture the internal variability at fixed $\theta$, each with a $10$ MTU spinup discarded and a $100$ MTU record retained. Windows of length $T$ at stride $\Delta t$ are cut from the stored records, so every configuration of Section~\ref{sec:results} uses the same trajectories.
}
\subsubsection{Lorenz-96 experiment}

Our second test case is the two-scale Lorenz-96 (L96) system \cite{lorenz1996proceedings, balwada2024learning}, \rev{a common testbed for subgrid parameterization, in which each slow variable is forced by the fast variables attached to it.}
The model consists of \(K\) slow variables \(X_k\) coupled to \(J\) fast variables
\(Y_{j,k}\) at each slow location and is governed by
\begin{align}
\dot{X}_k &= X_{k-1}(X_{k+1}-X_{k-2}) - X_k + F
            - \frac{hc}{b} \sum_{j=1}^{J} Y_{j,k}, \nonumber\\
\dot{Y}_{j,k} &= cb\,Y_{j+1,k}(Y_{j-1,k}-Y_{j+2,k})
                - c\,Y_{j,k} + \frac{hc}{b} X_k,
\label{eq:l96}
\end{align}
with cyclic indices in \(k\) and \(j\).
Throughout, we fix the system size to \(K=8\) slow variables and \(J=8\) fast
variables.

The parameter vector is
\[
\theta = (F,h,c,b),
\]
where \(F\) controls the large-scale forcing, \(h\) is the coupling strength between
slow and fast variables, \(c\) is the time-scale separation, and \(b\) is the amplitude
scaling of the fast variables.
\rev{The four parameters act through different variables and on different time scales, which is why the choice of what to observe matters here.}

\rev{
Parameters are drawn independently from Gaussians with means $(F,h,c,b) = (13, 1, 10, 10)$ and standard deviations $(1.5, 0.5, 2, 1)$, truncated by rejection to $F \in [10,16]$, $h \in [0.5,2]$, $c \in [5,15]$ and $b \in [8,12]$. As for Lorenz-63, $1024$ parameter values are drawn, fifty trajectories are integrated per value from distinct initial conditions, a $10$ MTU spinup is discarded and a $100$ MTU record is retained.

\label{sec:l96setup}
Observations may constrain the combined effect of several parameters more
strongly than the parameters individually. We  assess $hc/b$, the
coefficient of the fast-field sum in the slow equation, and the product $c\,b$,
as well as the four individual parameters. Each combination is formed from
every posterior draw before its mean is scored. 

\subsection{The experimental program}
Table~\ref{tab:axes} summarizes observation and conditioning choices. All configurations reuse stored simulations, while each requires separate training with three seeds. Within a trained setting, new observations require no further integration. This separates simulation reuse across designs from amortized inference across observations.

}
\begin{table}[ht]
\color{black}
\centering\small
\begin{tabular}{@{}ll@{}}
\toprule
axis & levels \\
\midrule
system & Lorenz-63; two-scale Lorenz-96 \\
observed variables & $(x,y,z)$ for L63; $X$, $Y$ or $(X,\Yb)$ for L96 \\
conditioning & trajectory; second-moment summaries; time mean; block-reduced; hybrid \\
window & $0.05$ to $100$ MTU \\
sampling interval & $\Delta t \in \{1, 5, 20, 50\}$ steps ($0.01$ to $0.5$ MTU) \\
observation noise & $0$, $5\%$, $20\%$ of each channel's standard deviation \\
\bottomrule
\end{tabular}
\caption{The axes of the experimental program and their levels. Each
configuration is one combination of a system, an observed variable set, a
conditioning type, a window, a sampling interval and a noise level, and every
configuration is trained with three seeds. Windows are given in model time
units (MTU); the step count $T$ and the stride $\Delta t$ are in units of
$0.01$ MTU for both systems, so $T = 200$ is $2$ MTU and $T = 10000$ is $100$
MTU. $\Yb_k$ denotes the average of the eight fast variables attached to slow site $k$.}
\label{tab:axes}
\end{table}
\section{Results}
\label{sec:results}
\color{black}
The experiments address three connected questions. Can we learn useful
summaries instead of specifying them in advance? How does their effectiveness
depend on record length and sampling interval? Which observed variables
provide information about the parameters of interest? Lorenz-63 and two-scale
Lorenz-96 let us examine these choices in controlled settings. Throughout, we
distinguish the information available in a record from how reliably a trained
estimator extracts it. Section~\ref{sec:ces} then examines posterior uncertainty
alongside parameter-estimation skill.

\subsection{Learning informative summaries}
\label{sec:l63}
We first compare learned and prescribed summaries of the same observation record. In the trajectory-conditioned approach, an encoder learns a summary directly from the trajectory as part of training the flow for parameter inference. In the summary-conditioned approach, the flow receives statistics chosen in advance. We use the same flow architecture, parameter split and training protocol in both cases to assess how well each representation supports inference from the same observations.

For Lorenz-96, both approaches start from the same $2$ MTU record of the slow variables and fast-field site averages, $(X,\Yb)$. Following Ref.~\cite{Cleary2021_CES_JCP}, we construct the benchmark by averaging each of $X$, $\Yb$, $X^2$, $X\Yb$ and $\Yb^2$ over all sites and times in the record. These statistics describe overall variability and covariation but discard temporal ordering and the identity of individual sites.
Trajectory conditioning improves recovery of all four parameters. For example, $R^2_c$ increases from $0.30$ with the prescribed summaries to $0.72$ with the learned summary (Table~\ref{tab:l96_fixedcost}). This comparison does not tell us how much of the gain comes from retaining temporal or spatial structure, or from differences in how accurately the two flows approximate their respective conditional posteriors.

\begin{table}[ht]
\color{black}
\centering\small
\begin{tabular}{@{}lccrrrr@{}}
\toprule
conditioning & window & $\Delta t$ (MTU) & $F$ & $h$ & $c$ & $b$ \\
\midrule
learned (trajectory) & $2$ MTU   & $0.01$  & $0.935$ & $0.904$ & $0.723$ & $0.759$ \\
\addlinespace
prescribed summaries & $2$ MTU   & $0.01$  & $0.759$ & $0.695$ & $0.301$ & $0.595$ \\
prescribed summaries & $10$ MTU  & $0.01$  & $0.870$ & $0.832$ & $0.465$ & $0.771$ \\
prescribed summaries & $50$ MTU  & $0.01$  & $0.933$ & $0.929$ & $0.708$ & $0.906$ \\
\addlinespace
prescribed summaries & $100$ MTU & $0.2$ & $0.937$ & $0.878$ & $0.690$ & $0.821$ \\
\bottomrule
\end{tabular}
\caption{Lorenz-96, learned and prescribed summaries of $(X,\Yb)$.
All entries are $R^2$ of the in-support posterior mean, averaged over three
training seeds. The trajectory and summary flows use the same $2$ MTU record
in the first two rows. Prescribed summaries are the site-and-window averages
of $X$, $\Yb$, $X^2$, $X\Yb$ and $\Yb^2$. Longer summary windows assess how
much coverage is needed to approach the short-trajectory skill. Sampling
intervals are in MTU; seed spreads are below $0.07$.
The full window sweep is in Table~\ref{tab:l96_fixedcost_full}.}
\label{tab:l96_fixedcost}
\end{table}

We next ask how much longer a record the prescribed summaries need to approach the skill of the learned summary from $2$ MTU. At the same sampling interval of $0.01$ MTU, prescribed summaries from a $50$ MTU record give comparable $R^2$ for $F$, $h$ and $c$. For $b$, a $10$ MTU record already gives comparable skill.
Learning the summary can  reduce the record length needed for useful inference,
but the benefit is specific to the parameter and the benchmark summaries.

We find that trajectory-based inference can be sensitive to the training seed. For Lorenz-63, means, standard deviations and correlations computed from a $2$ MTU record give $R^2_\sigma=0.93$. Using the same record, the trajectory-conditioned model with the lowest validation loss reaches $0.97$, but the mean score across three training seeds is only $0.30$ (Table~\ref{tab:l63_learning}). The prescribed summaries show that the record contains useful information about $\sigma$, even though some training runs fail to use it effectively.
Additional tests show that giving the flow both the learned representation and prescribed summaries of the same trajectory makes performance more consistent across training seeds, allowing the encoder to learn features beyond those captured by the prescribed summaries. Full comparisons and seed sensitivity are reported in  Appendices~\ref{sec:app_conditioning} and~\ref{sec:app_seeds}.

\begin{table}[ht]
\color{black}
\centering\small
\begin{tabular}{@{}llcrrr@{}}
\toprule
conditioning & window & $\Delta t$ (MTU) & $\rho$ & $\sigma$ & $\beta$ \\
\midrule
mean, sd and correlation & $2$ MTU & $0.01$ & $0.969$ & $0.926$ & $0.847$ \\
trajectory & $2$ MTU & $0.01$ & $0.964$ & $0.297$ & $0.315$ \\
\addlinespace
mean, sd and correlation & $10$ MTU & $0.2$ & $0.975$ & $0.490$ & $0.938$ \\
hybrid & $10$ MTU & $0.2$ & $0.990$ & $0.959$ & $0.988$ \\
\bottomrule
\end{tabular}
\caption{Lorenz-63, extracting information from the same observed record.
Entries are three-seed-mean $R^2$ of in-support posterior means. Each pair
uses identical samples. The hybrid combines the trajectory with its own
means, standard deviations and pairwise correlations. Raw-trajectory training
at $2$ MTU is unstable; the validation-selected model reaches $R^2_\sigma=0.967$
and $R^2_\beta=0.986$. Full conditioning comparisons and selected-model scores
are in Tables~\ref{tab:l63_summary}, \ref{tab:l63_trajectory}
and~\ref{tab:l63_trajectory_seeds}.}
\label{tab:l63_learning}
\end{table}

\subsection{Temporal coverage and sampling interval}
\label{sec:results_temporal}
Longer records can sample a wider range of states, while finer sampling can resolve faster changes. To examine these effects, we first vary the sampling interval at fixed record length and then extend the coarsely sampled record. We train a separate estimator for each observation design.
For a $2$ MTU Lorenz-96 record, increasing the sampling interval from $0.01$ to $0.2$ MTU does not weaken recovery of $F$ and $h$ from $X$. The parameters $c$ and $b$ remain weakly constrained at both intervals. With $Y$ as input, the same change reduces $R^2_c$ from $0.88$ to $0.23$ and $R^2_b$ from $0.79$ to $0.45$ (Table~\ref{tab:l96_sampling}). Recovery from $(X,\Yb)$ also weakens.

\begin{table}[ht]
\color{black}
\centering\small
\begin{tabular}{@{}llcrrrr@{}}
\toprule
observed & window & $\Delta t$ (MTU) & $F$ & $h$ & $c$ & $b$ \\
\midrule
$X$ & $2$ MTU  & $0.01$  & $0.888$ & $0.614$ & $0.124$ & $0.003$ \\
$X$ & $2$ MTU  & $0.2$ & $0.945$ & $0.679$ & $0.127$ & $0.057$ \\
\addlinespace
$Y$ & $2$ MTU  & $0.01$  & $0.589$ & $0.929$ & $0.884$ & $0.786$ \\
$Y$ & $2$ MTU  & $0.2$ & $0.384$ & $0.840$ & $0.226$ & $0.449$ \\
$Y$ & $20$ MTU & $0.2$ & $0.829$ & $0.966$ & $0.795$ & $0.901$ \\
\addlinespace
$(X,\Yb)$ & $2$ MTU & $0.01$  & $0.935$ & $0.904$ & $0.723$ & $0.759$ \\
$(X,\Yb)$ & $2$ MTU & $0.2$ & $0.931$ & $0.815$ & $0.516$ & $0.476$ \\
\bottomrule
\end{tabular}
\caption{
Effect of record length and sampling interval on inference from Lorenz-96 trajectories. Each score is the mean $R^2$ across three training seeds, with a spread below $0.07$. For a $2$ MTU record, increasing the sampling interval from $0.01$ to $0.2$ MTU reduces the number of time points retained per trajectory from $200$ to $10$. At the same interval of $0.2$ MTU, a $20$ MTU $Y$ record contains $100$ time points. Results for intermediate sampling intervals are reported in Table~\ref{tab:l96_sampling_full}.}
\label{tab:l96_sampling}
\end{table}

Extending the $Y$ record to $20$ MTU while keeping the sampling interval at $0.2$ MTU raises $R^2_c$ to $0.80$ and $R^2_b$ to $0.90$. The longer record contains $100$ samples, half as many as the finely sampled $2$ MTU record. Longer coverage can  compensate for coarser sampling even with fewer samples. The comparison does not reveal which features of the longer record the encoder uses to recover that skill.

Finer sampling does not always improve inference with the trained estimator. The modest gains from coarsening $X$ cannot reflect added information, because the finely sampled record contains all the coarse samples. One possibility is that the encoder, whose capacity is fixed, can extract the relevant features more readily from a shorter sequence.
Lorenz-63 also shows that a small number of samples can be sufficient for useful inference. With just four samples spaced $0.5$ MTU apart, the mean $R^2_\sigma$ is close to $0.89$, with consistent performance across seeds (Table~\ref{tab:l63_sampling_seeds}).

\subsection{Choosing observed variables}
\label{sec:l96}
We next examine how the choice of input variables affects parameter recovery. For Lorenz-96, we compare records of $X$ alone, $Y$ alone, or $(X,\Yb)$ (Table~\ref{tab:l96_observability}). These inputs give the estimator access to the slow variables, the fast variables, or the slow variables together with site averages of the fast field.

From $X$ alone, $c$ and $b$ remain weakly constrained over the tested $2$--$100$ MTU windows. Adding eight fast-field site averages at $2$ MTU raises $R^2_c$ from $0.12$ to $0.72$ and $R^2_b$ from approximately zero to $0.76$. For these parameters, retaining a fast-process diagnostic is more effective than extending the slow-variable record. Using $Y$ alone also improves their recovery, but gives weaker recovery of $F$ than $X$ does.

\begin{table}[ht]
\color{black}
\centering\small
\begin{tabular}{@{}lcrrrrrr@{}}
\toprule
observed & window & $F$ & $h$ & $c$ & $b$ & $hc/b$ & $c\,b$ \\
\midrule
$X$       & $2$ MTU   & $0.888$ & $0.614$ & $0.124$ & $0.003$ & $0.847$ & $0.026$ \\
$X$       & $100$ MTU & $0.957$ & $0.714$ & $0.132$ & $0.077$ & $0.941$ & $0.022$ \\
$Y$       & $2$ MTU   & $0.589$ & $0.929$ & $0.884$ & $0.786$ & $0.956$ & $0.921$ \\
$(X,\Yb)$ & $2$ MTU   & $0.935$ & $0.904$ & $0.723$ & $0.759$ & $0.978$ & $0.870$ \\
\bottomrule
\end{tabular}
\caption{Lorenz-96, the effect of observed variables
on trajectory-conditioned inference. Entries are three-seed-mean $R^2$ of
in-support posterior means, with seed spreads below $0.06$. The $2$ MTU records
are sampled every $0.01$ MTU and the $100$ MTU record every $0.2$ MTU. Composite
means are computed by transforming each posterior draw. Adding $\Yb$ improves
recovery of individual fast-process parameters; slow-only observations chiefly
constrain their combined influence. }
\label{tab:l96_observability}
\end{table}

\begin{figure}[ht]
\color{black}
\centering
\includegraphics[width=0.66\textwidth]{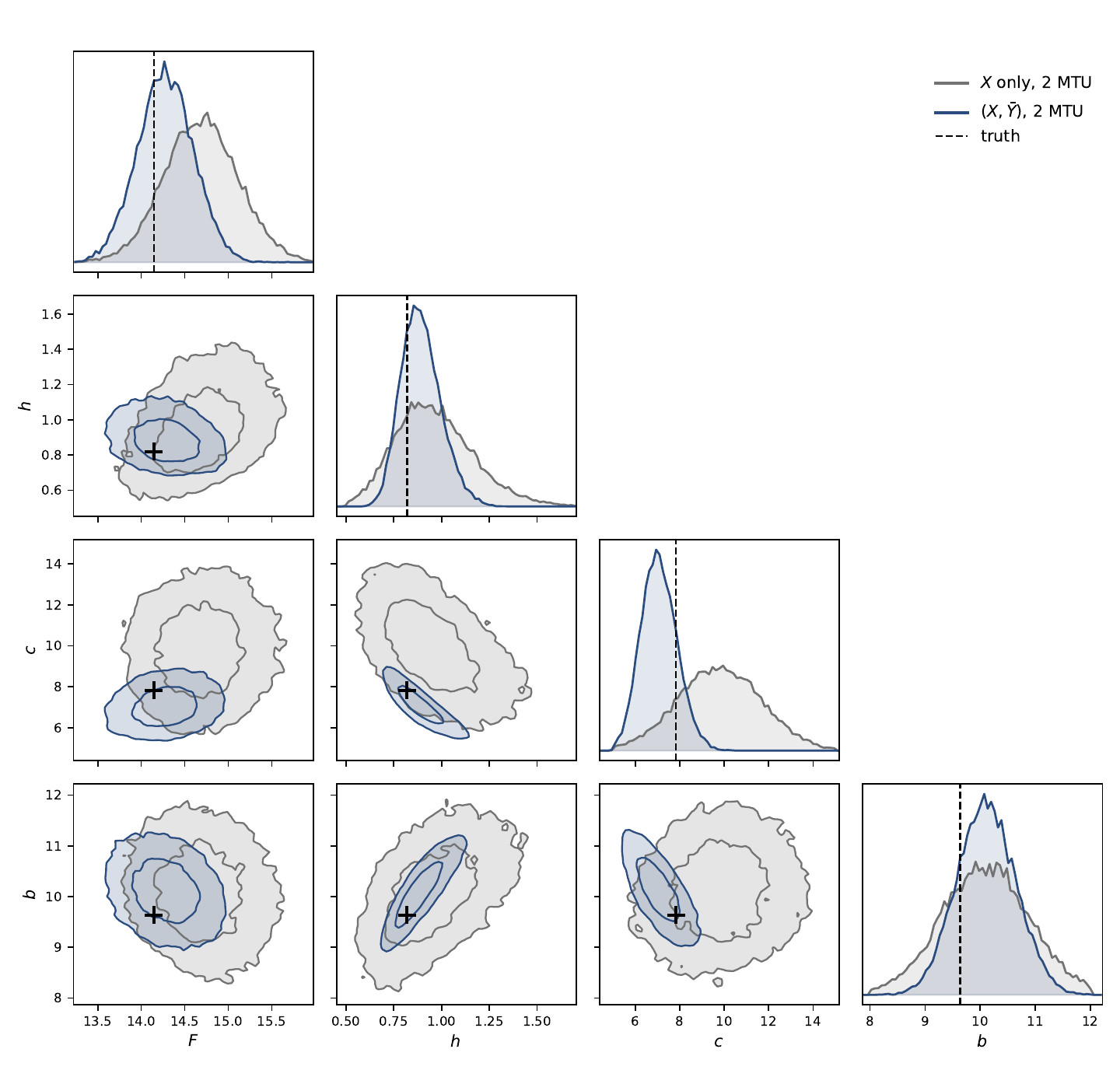}
\caption{Lorenz-96, posterior
constraints from $X$ alone (grey) and $(X,\Yb)$ (blue) for the same held-out
$2$ MTU record, sampled every $0.01$ MTU. The truth is
$(F,h,c,b)=(14.14,0.82,7.83,9.64)$. Diagonal panels show marginal densities;
lower panels show pairwise $50\%$ and $90\%$ highest-density regions. Dashed
lines and crosses mark the truth. Adding fast-field averages concentrates the
$c$ and $b$ marginals and shortens the $h$--$b$ ridge. This is an illustrative
observation; aggregate estimation skill is reported in
Table~\ref{tab:l96_observability}.}
\label{fig:l96_pairplot}
\end{figure}

The posterior also helps explain what is learned from slow observations.
Although individual $c$ and $b$ are weakly recovered, the composite $hc/b$
is estimated with high skill. Slow-only observations can constrain the net
influence of the fast process without tightly separating its parameters.
Figure~\ref{fig:l96_pairplot} illustrates broad marginals and an $h$--$b$ ridge,
which becomes more concentrated after $\Yb$ is added.
The weak recovery of $c$ and $b$ from $X$ alone persists across the tested conditioning choices and training seeds. This points to weak practical identifiability in these settings, although it does not rule out stronger constraints from other estimators or longer records.

\section{Posterior quality and amortized inference}
\label{sec:ces}
Accurate posterior means are only part of the inference task. We also want to
know whether the reported uncertainty gives useful parameter constraints.
We  compare flows with Calibrate--Emulate--Sample (CES)
\cite{Cleary2021_CES_JCP}, which calibrates an ensemble against prescribed
statistics, fits Gaussian-process emulators and samples a posterior using
MCMC. Both methods use the same priors and observed realizations. The full
protocol is in Appendix~\ref{sec:si_ces}.

For each system, we select eight distinct parameter values from the flow's held-out test set and randomly choose one trajectory for each. We keep these trajectories fixed across comparisons and calibrate CES separately for each observation. CES uses records of $2$, $20$ and $100$ MTU for Lorenz-96 and $10$ MTU for Lorenz-63.
For each parameter, we average the posterior standard deviation and the log posterior density at its true value across the eight observations. Coverage is the fraction of observations for which the central $90\%$ interval contains the true value. With only eight observations, coverage changes in steps of $1/8$ and provides a limited assessment of calibration.

\paragraph{Distinguishing representation and inference.}
We first give both methods identical summary statistics to compare their posterior estimates with the input held fixed (Appendix~\ref{sec:app_ces_extra}). On Lorenz-96, the flow has higher average marginal log scores at $2$ MTU, although several differences are small. At $20$ MTU, which method performs better depends on the parameter. At $100$ MTU, CES has higher log scores for all four parameters, with almost no difference for $F$ (Table~\ref{tab:ces_matched}). Its marginal posteriors are generally as wide as or wider than those of the flow. This illustrates why a narrower posterior is not necessarily better and why we also assess the density assigned to the true parameter value.

Together with the flow comparisons in Section~\ref{sec:l63}, these tests suggest that both the choice of summary and the posterior-estimation method affect performance on short records. We also compare the flow with an emulator-based sampler 1530 held-out trajectories  to assess accuracy and coverage. In this comparison, one emulator is trained per configuration and reused across observations (Appendix~\ref{sec:app_engines}).

\begin{figure}[ht]
\color{black}
\centering
\includegraphics[width=0.78\textwidth]{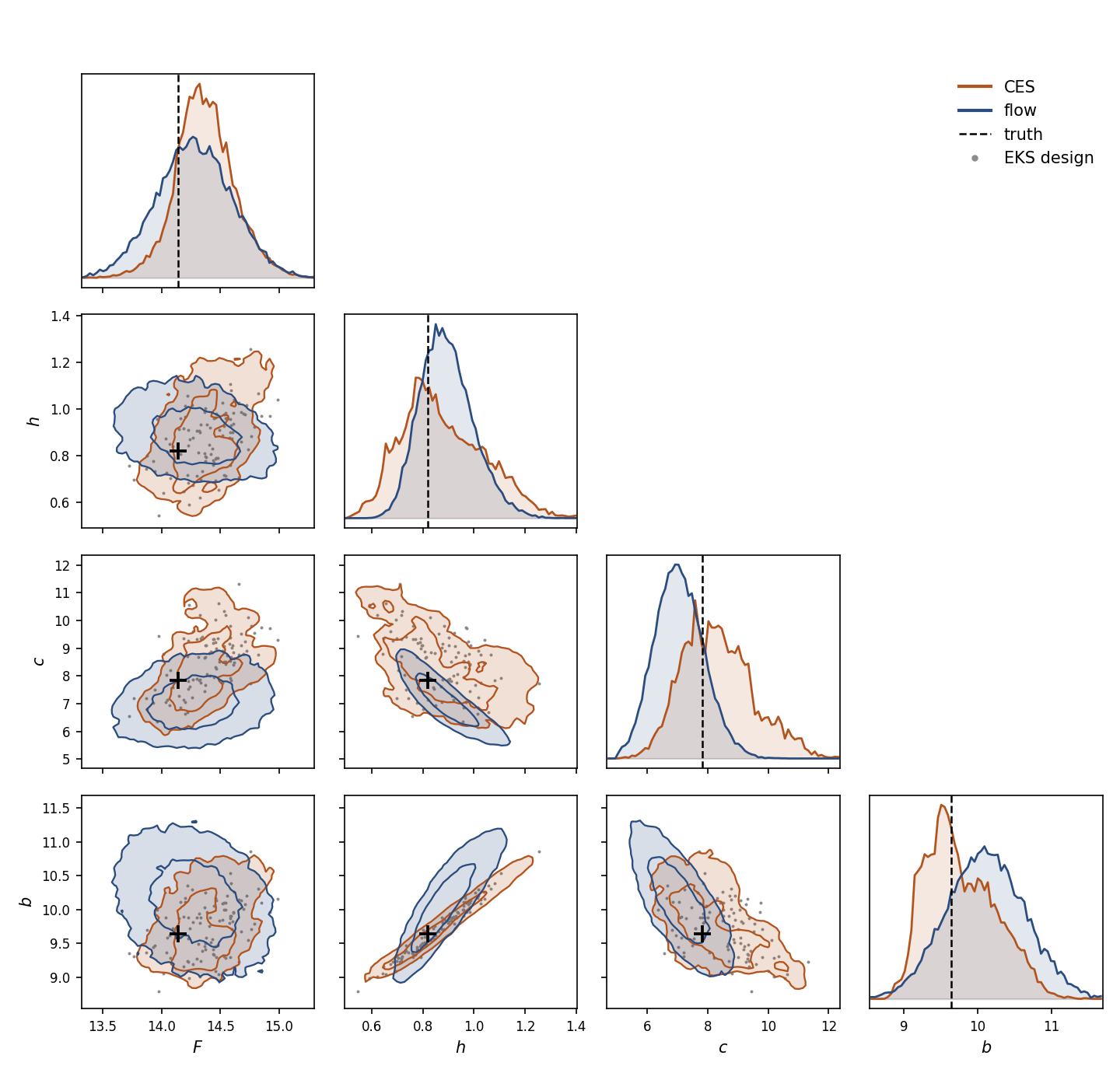}
\caption{Lorenz-96, short-trajectory and
long-statistics posteriors for one held-out observation with
$(F,h,c,b)=(14.14,0.82,7.83,9.64)$. CES (orange) uses $100$ MTU of statistics
sampled every $0.2$ MTU; the flow (blue) uses the first $2$ MTU of $(X,\Yb)$
sampled every $0.01$ MTU. Diagonal panels show marginal densities and lower
panels show $50\%$ and $90\%$ highest-density regions. Dashed lines and crosses
mark the truth; grey dots are the final ensemble Kalman sampler population.
Both methods capture an $h$--$b$ ridge, with different concentration along and
across it. Posterior quality is reported in
Table~\ref{tab:ces_perobs}.}
\label{fig:ces_perobs}
\end{figure}

\paragraph{Posterior constraints from short records.}
On Lorenz-96, the flow conditioned on a $2$ MTU trajectory of $(X,\Yb)$ gives posterior standard deviations comparable to CES using statistics from a $100$ MTU record (Table~\ref{tab:ces_perobs}). Average log scores favor the flow for $h$ and $c$ and CES for $F$ and $b$. Both methods capture correlations between parameters (Figure~\ref{fig:ces_perobs}); results for composite parameters are reported in Table~\ref{tab:ces_perobscomp}.
One qualification is that the longer CES records are sampled every $0.2$ MTU. Denser sampling could improve that baseline, as suggested by the summary-conditioned flow results in Table~\ref{tab:l96_fixedcost_full}.

\begin{table}[ht]
\color{black}
\centering\small
\begin{tabular}{@{}lcccc@{}}
\toprule
 & $F$ & $h$ & $c$ & $b$ \\
\midrule
posterior sd, CES with $100$ MTU of statistics & $0.335$ & $0.130$ & $0.939$ & $0.354$ \\
posterior sd, flow with $2$ MTU of trajectory  & $0.269$ & $0.104$ & $0.772$ & $0.384$ \\
\addlinespace
coverage $90\%$, CES  & $1.00$ & $0.88$ & $0.88$ & $0.75$ \\
coverage $90\%$, flow & $0.88$ & $0.88$ & $0.88$ & $1.00$ \\
\addlinespace
log score, CES  & $-0.09$ & $0.67$ & $-1.35$ & $-0.35$ \\
log score, flow & $-0.28$ & $1.01$ & $-1.22$ & $-0.39$ \\
\bottomrule
\end{tabular}
\caption{Lorenz-96, posterior quality for eight
held-out observations. CES uses the five statistics of $100$ MTU sampled every
$0.2$ MTU; the flow uses the first $2$ MTU of the same $(X,\Yb)$ record sampled
every $0.01$ MTU.  Standard
deviations and marginal log densities at the truth are averaged over
observations; coverage is the fraction of central $90\%$ intervals containing
the truth. Methods share priors but differ in record length, sampling,
representation and inference procedure. Matched-statistics controls are in
Table~\ref{tab:ces_matched}.}
\label{tab:ces_perobs}
\end{table}

On Lorenz-63, prescribed summaries from a short record also provide useful constraints (Table~\ref{tab:ces_perobs63}). For $\sigma$, the $2$ MTU trajectory flow gives a posterior standard deviation of $0.30$ and log score of $0.11$, compared with $0.88$ and $-0.87$ for CES using $10$ MTU of raw moments. A flow conditioned on means, standard deviations and correlations from the same $2$ MTU record gives a standard deviation of $0.27$ and log score of $0.25$. Much of the improvement over CES is therefore already available through prescribed short-window summaries.
Hybrid conditioning gives the highest $\sigma$ log score among these configurations. Some raw-trajectory models, however, have low empirical coverage despite high posterior-mean skill.

\paragraph{Cost and reuse.}
Once trained, the flow returns a posterior in about twenty milliseconds.
In our implementation, one $100$ MTU Lorenz-96 CES posterior requires $2000$
calibration runs and about fifty minutes of integration, emulator fitting and
sampling.

The initial investment in the flow remains important. Training one $2$ MTU
$(X,\Yb)$ flow took sixteen minutes on four CPU cores. Including spinup,
training-record integration is comparable to roughly two $100$ MTU CES
observations, under accounting that excludes validation and test simulations,
flow training and model selection (Appendix~\ref{sec:app_ces_extra}).
Simulations can be reused to compare observation designs, although each
configuration here requires separate training. The resulting estimators
support repeated inference for compatible records. Changes to dynamics,
priors or observation settings may require further simulations and retraining.

\clearpage
\section{Conclusions}
Parameter estimation often relies on prescribed summary statistics. Choosing them requires knowing which features of the observations constrain the parameters, which can be difficult in models with interacting processes and time scales. Longer records can sample a wider range of states and improve estimates of stationary statistics, but spatial or temporal averaging may still obscure features useful for inference. We examined this challenge using Lorenz-63 and two-scale Lorenz-96 as controlled test cases.

Our framework uses the encoder's representation of each trajectory as its summary statistic. We train the encoder jointly with a conditional normalizing flow to estimate the parameter posterior, encouraging the summary to retain information useful for inference. What the summary can retain depends on the observed variables, record length and sampling interval.
Our experiments show that learned summaries can match the estimation skill of prescribed statistics for several parameters using substantially shorter records. The value of a record also depends on which parameters we want to estimate. In Lorenz-96, for example, adding fast-field site averages to a short record improved recovery of the fast-process parameters more than extending the slow-variable record.

Together, these findings connect summary learning with observation design. The observed variables, window length, and sampling interval determine the information available to an estimator and should be assessed jointly. A longer record may compensate for coarse sampling, yet provide limited improvement when the observed variables weakly constrain the parameters of interest. Learning can identify useful features within a chosen record, but cannot recover information absent from it. Comparisons across observation settings can  inform which diagnostics to archive, how frequently to sample them, and how long to retain them. Such assessments should consider posterior calibration alongside parameter-estimation skill and posterior concentration.

Amortization makes repeated assessment practical once a suitable simulation ensemble is available. Existing perturbed-parameter ensembles with archived trajectories can support comparisons of observation settings, with simulations reused across configurations. Each configuration requires training, but a trained estimator returns posteriors for compatible new records without further model integration. This investment can support repeated calibration and observation-design analyses. Application to climate observations or reanalysis additionally requires comparable simulated and observed quantities and appropriate treatment of observational uncertainty and model discrepancy.

Several limitations remain. Joint training encourages informative summaries without guaranteeing sufficiency or optimality, and weak inference may reflect either limited observational information or failure to extract information that is present. Training requires adequate coverage of parameter space and internal variability, and the inferred posterior remains tied to the training prior and observation setting. Our experiments involve few parameters and no model discrepancy. Extending the framework to climate models will require scalable estimators, representative simulation ensembles, and reliable uncertainty assessment so that comparisons of observation designs reflect parameter information rather than limitations of the estimator.

\color{black}

\section*{Code availability statement}
Our code is available in the \href{https://github.com/Sshamekh/posterior-from-trajectories}{\rev{posterior-from-trajectories}} repository.
\begin{acknowledgments}
We acknowledge funding from NSF through the Learning the Earth with Artificial Intelligence and Physics (LEAP) Science and Technology Center (STC) (Award \#2019625). VA acknowledges support from a PIVOT Research award (Award \#12871) from the Simons Foundation.
\end{acknowledgments}

\bibliography{references}

\newpage
\appendix
\color{black}
\counterwithin{table}{section}
\renewcommand{\thetable}{\Alph{section}\arabic{table}}
\counterwithin{figure}{section}
\renewcommand{\thefigure}{\Alph{section}\arabic{figure}}
\section{Additional conditioning and temporal-design comparisons}
\label{sec:app_conditioning}
This appendix gives the full comparison of conditioning inputs for Lorenz-63, along with an example posterior. When variances are nonzero, means, standard deviations and correlations contain the same information as means and raw second moments. The choice of how to express these statistics can still affect approximate inference (Appendix~\ref{sec:app_engines}). For Lorenz-96, the tables give the full results across the tested record lengths and sampling intervals.

\begin{table}[ht]
\color{black}
\centering\small
\begin{tabular}{@{}llccrrr@{}}
\toprule
conditioning & window & $T$ & $\Delta t$ & $\rho$ & $\sigma$ & $\beta$ \\
\midrule
time mean of $(x,y,z)$ & $2$ MTU   & $200$   & $1$  & $0.762$ & $0.000$  & $0.081$ \\
                       & $10$ MTU  & $1000$  & $20$ & $0.923$ & $-0.018$ & $0.053$ \\
                       & $100$ MTU & $10000$ & $20$ & $0.936$ & $-0.014$ & $0.017$ \\
\addlinespace
mean, sd and correlation & $0.05$ MTU & $5$    & $1$  & $0.932$ & $0.895$ & $0.909$ \\
                         & $0.5$ MTU  & $50$   & $1$  & $0.934$ & $0.937$ & $0.833$ \\
                         & $2$ MTU    & $200$  & $1$  & $0.969$ & $0.926$ & $0.847$ \\
                         & $10$ MTU   & $1000$ & $20$ & $0.975$ & $0.490$ & $0.938$ \\
\addlinespace
raw first and second moments & $1$ MTU   & $100$   & $1$  & $0.945$ & $0.686$ & $0.738$ \\
                             & $5$ MTU   & $500$   & $1$  & $0.980$ & $0.773$ & $0.933$ \\
                             & $10$ MTU  & $1000$  & $1$  & $0.978$ & $0.833$ & $0.977$ \\
                             & $10$ MTU  & $1000$  & $20$ & $0.988$ & $0.384$ & $0.940$ \\
                             & $100$ MTU & $10000$ & $20$ & $0.985$ & $0.824$ & $0.992$ \\
\bottomrule
\end{tabular}
\caption{Lorenz-63 parameter inference from prescribed summaries. For each parameter, we compute $R^2$ between posterior means and true values on the held-out test set and average the scores across three training seeds. Posterior means use only draws within the prior bounds. We compare three sets of statistics computed over each window. The first contains the three time means. The other two contain nine statistics each, either the means, standard deviations and pairwise correlations or the means and six raw second moments, such as $\langle x^2\rangle$ and $\langle xy\rangle$. Second-moment summaries recover $\rho$ and $\beta$ from long windows. Recovery of $\sigma$ depends on the sampling interval and, at coarse sampling, on the window length. The spread across seeds is below $0.05$ except for $\beta$ with mean, standard deviation and correlation inputs, where it is $0.05$--$0.08$. }
\label{tab:l63_summary}
\end{table}

\begin{table}[ht]
\color{black}
\centering\small
\begin{tabular}{@{}llcccrrr@{}}
\toprule
conditioning & window & $T$ & $\Delta t$ & samples & $\rho$ & $\sigma$ & $\beta$ \\
\midrule
trajectory    & $0.05$ MTU & $5$     & $1$  & $5$   & $0.808$ & $0.476$ & $0.718$ \\
trajectory    & $0.5$ MTU  & $50$    & $1$  & $50$  & $0.979$ & $0.968$ & $0.974$ \\
trajectory    & $2$ MTU    & $200$   & $1$  & $200$ & $0.995$ & $0.967$ & $0.986$ \\
trajectory    & $10$ MTU   & $1000$  & $20$ & $50$  & $0.991$ & $0.923$ & $0.996$ \\
trajectory    & $100$ MTU  & $10000$ & $20$ & $500$ & $0.988$ & $0.957$ & $0.996$ \\
block-reduced & $100$ MTU  & $10000$ & $20$ & $25$  & $0.987$ & $0.897$ & $0.993$ \\
hybrid        & $10$ MTU   & $1000$  & $20$ & $50$  & $0.993$ & $0.987$ & $0.993$ \\
hybrid        & $100$ MTU  & $10000$ & $20$ & $500$ & $0.967$ & $0.911$ & $0.994$ \\
\addlinespace
trajectory    & $2$ MTU    & $200$   & $20$ & $10$  & $0.986$ & $0.937$ & $0.995$ \\
trajectory    & $2$ MTU    & $200$   & $50$ & $4$   & $0.982$ & $0.905$ & $0.981$ \\
\bottomrule
\end{tabular}
\caption{Lorenz-63, trajectory-conditioned inference as a function of window
length and sampling interval. $R^2$ on the held-out split for the flow
conditioned on the raw $(x,y,z)$ trajectory and, in three rows, on the
block-reduced trajectory (twenty-five blocks of per-variable mean and standard
deviation) or on the hybrid input (the trajectory together with the mean,
standard deviation and correlations of the same samples); the samples column
is the number of time points the flow sees. The upper block varies the window;
the lower block holds the window at $2$ MTU and coarsens the sampling. Each row
is the model selected by validation loss among three seeds. For the
raw-trajectory rows the seeds disagree on $\sigma$ and $\beta$, while the
block-reduced and hybrid rows are seed-stable; the three-seed means are in
Tables~\ref{tab:l63_trajectory_seeds} and~\ref{tab:l63_sampling_seeds}.}
\label{tab:l63_trajectory}
\end{table}

\begin{table}[ht]
\color{black}
\centering\small
\begin{tabular}{@{}llccrrrr@{}}
\toprule
observed & conditioning & window & $\Delta t$ & $F$ & $h$ & $c$ & $b$ \\
\midrule
$(X,\Yb)$ & trajectory & $2$ MTU   & $1$  & $0.935$ & $0.904$ & $0.723$ & $0.759$ \\
\addlinespace
$(X,\Yb)$ & second-moment summaries & $2$ MTU   & $1$  & $0.759$ & $0.695$ & $0.301$ & $0.595$ \\
$(X,\Yb)$ & second-moment summaries & $5$ MTU   & $1$  & $0.854$ & $0.781$ & $0.398$ & $0.712$ \\
$(X,\Yb)$ & second-moment summaries & $10$ MTU  & $1$  & $0.870$ & $0.832$ & $0.465$ & $0.771$ \\
$(X,\Yb)$ & second-moment summaries & $20$ MTU  & $1$  & $0.906$ & $0.872$ & $0.573$ & $0.831$ \\
$(X,\Yb)$ & second-moment summaries & $50$ MTU  & $1$  & $0.933$ & $0.929$ & $0.708$ & $0.906$ \\
\addlinespace
$(X,\Yb)$ & second-moment summaries & $20$ MTU  & $20$ & $0.882$ & $0.781$ & $0.479$ & $0.658$ \\
$(X,\Yb)$ & second-moment summaries & $100$ MTU & $20$ & $0.937$ & $0.878$ & $0.690$ & $0.821$ \\
\bottomrule
\end{tabular}
\caption{Lorenz-96, trajectory against summary statistics on the same record
and at the same skill. $R^2$, three-seed mean, all rows are flows. Row one is
conditioned on the raw trajectory of $(X,\Yb)$ over $2$ MTU. The remaining rows
are conditioned on the five second-moment summaries of the same variables, the
site-and-window averages of $X$, $\Yb$, $X^2$, $X\Yb$ and $\Yb^2$, computed
from the samples retained at the stated $\Delta t$, in steps of $0.01$ MTU.
Reading across the $2$ MTU rows compares what is retained from the same
record; reading down the summary rows gives the record length the summaries
need to reach the skill of the $2$ MTU trajectory, between $20$ and $50$ MTU
at $\Delta t = 1$ and between $20$ and $100$ MTU at $\Delta t = 20$. Seed
spreads are below $0.07$.}
\label{tab:l96_fixedcost_full}
\end{table}

\begin{table}[ht]
\color{black}
\centering\small
\begin{tabular}{@{}llcrrrr@{}}
\toprule
observed & window & $\Delta t$ (steps) & $F$ & $h$ & $c$ & $b$ \\
\midrule
$X$ & $2$ MTU  & $1$  & $0.888$ & $0.614$ & $0.124$ & $0.003$ \\
$X$ & $2$ MTU  & $5$  & $0.901$ & $0.660$ & $0.123$ & $0.058$ \\
$X$ & $2$ MTU  & $20$ & $0.945$ & $0.679$ & $0.127$ & $0.057$ \\
\addlinespace
$Y$ & $2$ MTU  & $1$  & $0.589$ & $0.929$ & $0.884$ & $0.786$ \\
$Y$ & $2$ MTU  & $5$  & $0.519$ & $0.918$ & $0.781$ & $0.742$ \\
$Y$ & $2$ MTU  & $20$ & $0.384$ & $0.840$ & $0.226$ & $0.449$ \\
$Y$ & $20$ MTU & $20$ & $0.829$ & $0.966$ & $0.795$ & $0.901$ \\
\addlinespace
$(X,\Yb)$ & $2$ MTU & $1$  & $0.935$ & $0.904$ & $0.723$ & $0.759$ \\
$(X,\Yb)$ & $2$ MTU & $5$  & $0.958$ & $0.920$ & $0.700$ & $0.738$ \\
$(X,\Yb)$ & $2$ MTU & $20$ & $0.931$ & $0.815$ & $0.516$ & $0.476$ \\
\bottomrule
\end{tabular}
\caption{Lorenz-96, effect of the sampling interval at a fixed window of $2$
MTU. $R^2$, three-seed mean, for flows conditioned on the raw trajectory of
the slow variables $X$, of the fast variables $Y$, or of $(X,\Yb)$, with the
number of samples in the window falling from $200$ to $10$ across the rows of
each block. The fourth $Y$ row extends the coarsely sampled window to $20$
MTU at the same $\Delta t$. Seed spreads are below $0.07$.}
\label{tab:l96_sampling_full}
\end{table}

\newpage
\section{Observation noise}

\label{sec:app_noise}
Independent Gaussian noise is added to trajectory samples at a stated fraction
of each channel's standard deviation. Hybrid summaries are computed from those
noisy samples. For summary-conditioned flows, noise was added either to the summaries at a
fraction of each summary's training-set standard deviation or to the samples
before forming summaries
(Table~\ref{tab:l63_noise}). Only the latter is a matched observational-noise
comparison with trajectory and hybrid inputs. Short-time changes can be
sensitive to sample noise, but training instability and the use of learned
representations prevent these tests from uniquely identifying the mechanism
of parameter recovery. On Lorenz-96, results from $X$ and $(X,\Yb)$ are largely
unchanged, with a modest decline for $c$ from $(X,\Yb)$ at $20\%$ noise
(Table~\ref{tab:l96_noise}). Noise robustness of $Y$-only inference was not tested.

\paragraph{Lorenz-63.}
Hybrid conditioning remains useful under observation noise (Table~\ref{tab:l63_noise}). At $10$ MTU, it retains $R^2_\sigma = 0.93$ under $5\%$ Gaussian noise and $0.77$ under $20\%$ noise, compared with $0.44$ and $0.31$ for summaries of the same noisy samples.
Raw-trajectory inference is more sensitive. For densely sampled $2$ MTU records, $5\%$ noise reduces $R^2_\sigma$ from $0.97$ to $0.01$ in the selected models. Sampling every $0.5$ MTU retains $0.61$ at this noise level, but only $0.09$ at $20\%$. Existing seed instability in the dense configuration limits interpretation of its decline. Hybrid conditioning provides the more reliable recovery across both noise levels.

\begin{table}[ht]
\color{black}
\centering\small
\begin{tabular}{@{}llrrrcrrr@{}}
\toprule
 & & \multicolumn{3}{c}{three-seed mean} & & \multicolumn{3}{c}{best seed} \\
\cmidrule(lr){3-5}\cmidrule(lr){7-9}
conditioning & noise & $\rho$ & $\sigma$ & $\beta$ & & $\rho$ & $\sigma$ & $\beta$ \\
\midrule
trajectory, $2$ MTU, $\Delta t = 1$ & $0$    & $0.964$ & $0.297$  & $0.315$ & & $0.995$ & $0.967$  & $0.986$ \\
                                    & $5\%$  & $0.974$ & $-0.019$ & $0.132$ & & $0.990$ & $0.014$  & $0.321$ \\
                                    & $20\%$ & $0.960$ & $-0.027$ & $0.035$ & & $0.980$ & $-0.035$ & $0.031$ \\
\addlinespace
trajectory, $2$ MTU, $\Delta t = 50$ & $0$    & $0.984$ & $0.885$ & $0.977$ & & $0.982$ & $0.905$ & $0.981$ \\
                                     & $5\%$  & $0.957$ & $0.591$ & $0.918$ & & $0.965$ & $0.610$ & $0.926$ \\
                                     & $20\%$ & $0.790$ & $0.065$ & $0.570$ & & $0.811$ & $0.092$ & $0.590$ \\
\addlinespace
hybrid, $10$ MTU, $\Delta t = 20$ & $0$    & $0.990$ & $0.959$ & $0.988$ & & $0.993$ & $0.987$ & $0.993$ \\
                                  & $5\%$  & $0.960$ & $0.768$ & $0.948$ & & $0.989$ & $0.932$ & $0.987$ \\
                                  & $20\%$ & $0.977$ & $0.749$ & $0.952$ & & $0.980$ & $0.769$ & $0.958$ \\
\addlinespace
statistics, $10$ MTU, noise on summaries & $0$    & $0.975$ & $0.490$ & $0.938$ & & $0.991$ & $0.500$ & $0.944$ \\
                                        & $5\%$  & $0.985$ & $0.494$ & $0.901$ & & $0.987$ & $0.502$ & $0.897$ \\
                                        & $20\%$ & $0.947$ & $0.385$ & $0.685$ & & $0.949$ & $0.395$ & $0.686$ \\
\addlinespace
statistics, $10$ MTU, noise on samples & $5\%$  & $0.965$ & $0.442$ & $0.906$ & & $0.979$ & $0.459$ & $0.916$ \\
                                        & $20\%$ & $0.967$ & $0.311$ & $0.779$ & & $0.967$ & $0.305$ & $0.774$ \\
\bottomrule
\end{tabular}
\caption{Lorenz-63, effect of observation noise. $R^2$ for the raw-trajectory
flow at $2$ MTU at two sampling intervals, for the hybrid flow at $10$ MTU, and
for the flow conditioned on the mean, standard deviation and correlations of
a $10$ MTU window sampled every $0.2$ MTU (statistics rows). For the trajectory and hybrid rows
the noise is added to every sample at the stated fraction of each channel's
standard deviation before any statistic is formed. For the summary flow the
table gives both conventions, noise added to the nine summaries at the stated
fraction of each summary's training-set standard deviation, and noise added
to the samples before the summaries are computed. Without noise only one of
the three $\Delta t = 1$ seeds recovers $\sigma$ and $\beta$; the $\Delta t =
50$ and hybrid rows are seed-stable.}
\label{tab:l63_noise}
\end{table}

\begin{table}[ht]
\color{black}
\centering\small
\begin{tabular}{@{}llrrrrrr@{}}
\toprule
observed & noise & $F$ & $h$ & $c$ & $b$ & $hc/b$ & $c\,b$ \\
\midrule
$X$ & $0$    & $0.888$ & $0.614$ & $0.124$ & $0.003$ & $0.847$ & $0.026$ \\
$X$ & $5\%$  & $0.881$ & $0.603$ & $0.101$ & $0.051$ & $0.842$ & $0.007$ \\
$X$ & $20\%$ & $0.888$ & $0.628$ & $0.111$ & $0.062$ & $0.853$ & $0.019$ \\
\addlinespace
$(X,\Yb)$ & $0$    & $0.935$ & $0.904$ & $0.723$ & $0.759$ & $0.978$ & $0.870$ \\
$(X,\Yb)$ & $5\%$  & $0.932$ & $0.912$ & $0.684$ & $0.750$ & $0.969$ & $0.828$ \\
$(X,\Yb)$ & $20\%$ & $0.925$ & $0.913$ & $0.669$ & $0.741$ & $0.959$ & $0.804$ \\
\bottomrule
\end{tabular}
\caption{Lorenz-96, effect of observation noise. $R^2$, three-seed mean, for
the flow conditioned on the trajectory of $X$ or of $(X,\Yb)$ over $2$ MTU at
$\Delta t = 1$, with Gaussian noise added to every sample at $0$, $5\%$ and
$20\%$ of each channel's standard deviation. Every entry is unchanged to
within seed spread except $c$ from $(X,\Yb)$, which loses $0.05$ at $20\%$.}
\label{tab:l96_noise}
\end{table}

\section{Seed dependence on Lorenz-63}
\label{sec:app_seeds}
Tables~\ref{tab:l63_trajectory_seeds} and~\ref{tab:l63_sampling_seeds} give the three-seed mean beside the model selected by validation loss for the configurations of Table~\ref{tab:l63_trajectory}. For the flows conditioned on the raw trajectory the seeds split on $\sigma$ and $\beta$, some above $0.9$ and the others near zero, so the three-seed mean does not represent any individual run. Every seed recovers $\rho$, and the hybrid and block-reduced rows are seed-stable.

\begin{table}[ht]
\color{black}
\centering\small
\begin{tabular}{@{}lccrrrcrrr@{}}
\toprule
 & & & \multicolumn{3}{c}{three-seed mean} & & \multicolumn{3}{c}{best seed} \\
\cmidrule(lr){4-6}\cmidrule(lr){8-10}
conditioning, window & $T$ & $\Delta t$ & $\rho$ & $\sigma$ & $\beta$ & & $\rho$ & $\sigma$ & $\beta$ \\
\midrule
trajectory, $0.05$ MTU & $5$     & $1$  & $0.677$ & $0.350$ & $0.301$ & & $0.808$ & $0.476$ & $0.718$ \\
trajectory, $0.5$ MTU  & $50$    & $1$  & $0.859$ & $0.556$ & $0.611$ & & $0.979$ & $0.968$ & $0.974$ \\
trajectory, $2$ MTU    & $200$   & $1$  & $0.964$ & $0.297$ & $0.315$ & & $0.995$ & $0.967$ & $0.986$ \\
trajectory, $10$ MTU   & $1000$  & $20$ & $0.988$ & $0.279$ & $0.784$ & & $0.991$ & $0.923$ & $0.996$ \\
trajectory, $100$ MTU  & $10000$ & $20$ & $0.976$ & $0.306$ & $0.447$ & & $0.988$ & $0.957$ & $0.996$ \\
\addlinespace
hybrid, $10$ MTU        & $1000$  & $20$ & $0.990$ & $0.959$ & $0.988$ & & $0.993$ & $0.987$ & $0.993$ \\
hybrid, $100$ MTU       & $10000$ & $20$ & $0.951$ & $0.877$ & $0.983$ & & $0.967$ & $0.911$ & $0.994$ \\
block-reduced, $100$ MTU & $10000$ & $20$ & $0.989$ & $0.890$ & $0.992$ & & $0.987$ & $0.897$ & $0.993$ \\
\bottomrule
\end{tabular}
\caption{Lorenz-63, trajectory-conditioned inference as a function of window
length. $R^2$ on the held-out split for the flow conditioned on
the raw $(x,y,z)$ trajectory (pooled encoder), on the trajectory together with
its own nine summary statistics (hybrid), and on the block-reduced trajectory
(twenty-five blocks of per-variable mean and standard deviation). Two scores
are given because the raw-trajectory flow trains to one of two solutions
depending on the seed. Some seeds recover $\sigma$ and $\beta$ with $R^2$
above $0.9$ and the others not at all. The best seed is the one with the
lowest validation loss, chosen without reference to the test split. The hybrid
and block-reduced rows are seed-stable (spread below $0.04$).}
\label{tab:l63_trajectory_seeds}
\end{table}

\begin{table}[ht]
\color{black}
\centering\small
\begin{tabular}{@{}lcrrrcrrr@{}}
\toprule
 & & \multicolumn{3}{c}{three-seed mean} & & \multicolumn{3}{c}{best seed} \\
\cmidrule(lr){3-5}\cmidrule(lr){7-9}
$\Delta t$ (steps) & samples & $\rho$ & $\sigma$ & $\beta$ & & $\rho$ & $\sigma$ & $\beta$ \\
\midrule
$1$  & $200$ & $0.964$ & $0.297$  & $0.315$ & & $0.995$ & $0.967$  & $0.986$ \\
$20$ & $10$  & $0.949$ & $0.296$  & $0.352$ & & $0.986$ & $0.937$  & $0.995$ \\
$50$ & $4$   & $0.984$ & $0.885$  & $0.977$ & & $0.982$ & $0.905$  & $0.981$ \\
\bottomrule
\end{tabular}
\caption{Lorenz-63, effect of the sampling interval at a fixed window of $2$
MTU ($T = 200$ steps). $R^2$ for the flow conditioned on the raw
trajectory, pooled encoder. The window is identical in every row; only the
number of samples inside it changes. The $\Delta t =
50$ row is the only raw-trajectory configuration in the sweep that is
seed-stable (spread below $0.03$).}
\label{tab:l63_sampling_seeds}
\end{table}

\section{Observability on Lorenz-96, full tables}
\label{sec:app_obs}

Tables~\ref{tab:l96_observability_full} and~\ref{tab:l96_composites} give the configurations observing $X$ alone, $Y$ alone or $(X,\Yb)$, with the two combinations of the fast-scale parameters scored from the same posterior draws by taking the combination of every draw. The block-reduced flow observing $X$ reduces a $20$ MTU record of a hundred samples to twenty-five blocks and gives similar $R^2$ for  $c$ and $b$ as the trajectory encoders at every sequence length from ten to five hundred samples, supporting weak practical identifiability across the tested estimators without establishing structural non-identifiability.

\begin{table}[ht]
\color{black}
\centering\small
\begin{tabular}{@{}llccrrrr@{}}
\toprule
observed & conditioning & window & $\Delta t$ & $F$ & $h$ & $c$ & $b$ \\
\midrule
$X$ & time mean          & $100$ MTU & $20$ & $0.779$ & $0.343$ & $0.013$ & $0.031$ \\
$X$ & mean and sd        & $2$ MTU   & $1$  & $0.542$ & $0.435$ & $0.035$ & $-0.034$ \\
$X$ & mean and sd        & $100$ MTU & $20$ & $0.859$ & $0.636$ & $0.123$ & $0.054$ \\
$X$ & block-reduced      & $20$ MTU  & $20$ & $0.926$ & $0.674$ & $0.122$ & $0.062$ \\
$X$ & trajectory         & $2$ MTU   & $1$  & $0.888$ & $0.614$ & $0.124$ & $0.003$ \\
$X$ & trajectory         & $20$ MTU  & $20$ & $0.962$ & $0.696$ & $0.124$ & $0.058$ \\
$X$ & trajectory         & $100$ MTU & $20$ & $0.957$ & $0.714$ & $0.132$ & $0.077$ \\
\addlinespace
$Y$ & mean and sd        & $2$ MTU   & $1$  & $0.254$ & $0.737$ & $0.342$ & $0.438$ \\
$Y$ & trajectory         & $2$ MTU   & $1$  & $0.589$ & $0.929$ & $0.884$ & $0.786$ \\
$Y$ & trajectory         & $20$ MTU  & $20$ & $0.829$ & $0.966$ & $0.795$ & $0.901$ \\
\addlinespace
$(X,\Yb)$ & trajectory   & $2$ MTU   & $1$  & $0.935$ & $0.904$ & $0.723$ & $0.759$ \\
$(X,\Yb)$ & trajectory   & $20$ MTU  & $20$ & $0.961$ & $0.899$ & $0.726$ & $0.799$ \\
$(X,\Yb)$ & trajectory   & $100$ MTU & $20$ & $0.968$ & $0.937$ & $0.801$ & $0.908$ \\
\bottomrule
\end{tabular}
\caption{Lorenz-96, identifiability as a function of the observed variables.
$R^2$ of the in-support posterior mean on the held-out split, three-seed mean,
for flows observing the slow variables $X$ only, the sixty-four fast variables
$Y$ only, or the slow variables together with the eight site averages of the
fast field $(X,\Yb)$. Conditioning is on the raw trajectory over the stated
window, on the block-reduced trajectory (twenty-five blocks of per-channel mean
and standard deviation), on the per-channel mean and standard deviation over
the window, or on the time mean alone. Seed spread is below $0.07$ on every row
except $(X,\Yb)$ at $100$ MTU, where it is $0.18$ on $c$ and the row 
gives the model selected by validation loss.}
\label{tab:l96_observability_full}
\end{table}

\begin{table}[ht]
\color{black}
\centering\small
\begin{tabular}{@{}llccrrrr@{}}
\toprule
observed & conditioning & window & $\Delta t$ & $c$ & $b$ & $hc/b$ & $c\,b$ \\
\midrule
$X$ & time mean     & $100$ MTU & $20$ & $0.013$ & $0.031$ & $0.482$ & $-0.028$ \\
$X$ & mean and sd   & $100$ MTU & $20$ & $0.123$ & $0.054$ & $0.867$ & $0.026$ \\
$X$ & block-reduced & $20$ MTU  & $20$ & $0.122$ & $0.062$ & $0.910$ & $0.015$ \\
$X$ & trajectory    & $2$ MTU   & $1$  & $0.124$ & $0.003$ & $0.847$ & $0.026$ \\
$X$ & trajectory    & $100$ MTU & $20$ & $0.132$ & $0.077$ & $0.941$ & $0.022$ \\
\addlinespace
$Y$ & mean and sd   & $2$ MTU   & $1$  & $0.342$ & $0.438$ & $0.698$ & $0.344$ \\
$Y$ & trajectory    & $2$ MTU   & $1$  & $0.884$ & $0.786$ & $0.956$ & $0.921$ \\
\addlinespace
$(X,\Yb)$ & trajectory & $2$ MTU   & $1$  & $0.723$ & $0.759$ & $0.978$ & $0.870$ \\
$(X,\Yb)$ & trajectory & $100$ MTU & $20$ & $0.801$ & $0.908$ & $0.985$ & $0.875$ \\
\bottomrule
\end{tabular}
\caption{Lorenz-96, the fast-scale parameters and their two combinations. $R^2$
of the composite coupling $hc/b$ that multiplies the fast field in the slow
equation, and of the product $c\,b$, computed from the
same posterior draws as Table~\ref{tab:l96_observability_full} by taking the
combination of every in-support draw and scoring its posterior mean against the
true combination. From the slow variable alone the composite is recovered while
$c$, $b$ and $c\,b$ are not; adding $\Yb$ substantially reduces the degeneracy in the tested settings. Seed spread on
the composites is below $0.04$; the $(X,\Yb)$ row at $100$ MTU gives the model
selected by validation loss, as in Table~\ref{tab:l96_observability_full}.}
\label{tab:l96_composites}
\end{table}

\section{Setup and scoring of the Calibrate--Emulate--Sample comparison}
\label{sec:si_ces}

\subsection{Configurations}
We run the full Calibrate--Emulate--Sample (CES) pipeline, including calibration, for four configurations. For Lorenz-96, we use records of $2$, $20$ and $100$ MTU, sampled every $0.01$ MTU for the shortest record and every $0.2$ MTU for the longer records. For canonical Lorenz-63, we use $10$ MTU records sampled every $0.01$ MTU.
Each configuration uses eight held-out observations, giving $32$ CES posteriors in the main comparison. Each is paired with one or more flow posteriors for the same observation. Both methods use a common simulation pipeline with the same integrator, numerical time step, spinup and prior.
An additional Lorenz-63 configuration trains the emulator using only the final calibration iteration, following the published prescription. This configuration is used only for the ablation in Table~\ref{tab:ces_ablation}.

\subsection{Observations}
We split the parameter values into disjoint training, validation and test sets in proportions of $70\%$, $15\%$ and $15\%$, using a fixed random seed. All realizations of a given parameter value remain in the same set. We select eight parameter values at evenly spaced positions in the sorted list of test-set indices. For each value, we randomly choose one stored trajectory and use it as the observation for both methods. The same eight trajectories are used across window lengths, always taking the first $T$ stored time steps.

CES receives each trajectory through a vector of summary statistics. For Lorenz-63, this contains the three time means and six raw second moments, $\langle x^2\rangle$, $\langle y^2\rangle$, $\langle z^2\rangle$, $\langle xy\rangle$, $\langle yz\rangle$ and $\langle zx\rangle$, giving nine statistics. For Lorenz-96, it contains averages of $X$, $\Yb$, $X^2$, $X\Yb$ and $\Yb^2$ over all sites and retained time points. Here, $\Yb_k$ is the mean of the eight fast variables at site $k$. All statistics use the stated sampling interval. At $\Delta t=20$, retained time points are $0.2$ MTU apart.

Each flow uses the same underlying trajectory with the window length, sampling interval and input type for which it was trained. The matched control receives exactly the same summary vector as CES. Trajectory-conditioned flows receive time series of the selected variables, with prescribed summaries added for hybrid inputs.

\subsection{Forward model and  internal variability}

CES uses the same fourth-order Runge--Kutta integrator as the dataset generation, with time steps of $0.002$ MTU for Lorenz-63 and $0.0005$ MTU for Lorenz-96. Each run starts from a random initial state and includes a $10$ MTU spinup. Each model evaluation returns the summary vector of one trajectory over the requested window, using time points retained every $0.01\,\Delta t$ MTU.

The covariance $\Gamma_{\mathrm{obs}}$ represents internal variability in the summary statistics and enters the CES likelihood as observation uncertainty. We estimate it from the sample covariance of summary vectors from $200$ independent trajectories at the prior mode. Each trajectory has its own spinup and uses the same window length and sampling interval as the observation. We estimate this covariance once per configuration and hold it fixed across the eight observations.
Matching the sampling interval matters because subsampling changes the variability of the statistics. For the Lorenz-96 fast-field statistics, the standard deviation at $\Delta t=20$ is $19$--$23\%$ higher than when all stored time points are used. The covariance must  describe the statistics computed from the time points actually retained in the observation.
\subsection{Calibration}

The calibration stage uses the ensemble Kalman sampler. It
starts from $M = 100$ particles drawn from the prior and runs for twenty
iterations. At each iteration the forward model is evaluated once at every
particle, giving $2000$ forward runs per observation. The update is the
implicit scheme of the CES paper. It combines a Kalman drift toward the observation using
the ensemble's own covariance, a drift toward the prior using the prior's
Gaussian moments, and injected noise scaled by the ensemble covariance, with
an adaptive step that shrinks as the ensemble's misfit to the observation grows
\cite{Cleary2021_CES_JCP}. Particles
are clipped to the prior's box after each update. The noise and the prior
drift are what stop the ensemble from collapsing to a point, so that its final
spread approximates the posterior's and the emulator is trained where the
posterior mass is. Every iteration's parameter values and summary
evaluations are kept.

\subsection{Emulation}

The emulator is one Gaussian process per summary component, mapping the
parameter vector to that component. Inputs and outputs are standardized. The
kernel is a constant amplitude times a Mat\'ern-$5/2$ with a separate length
scale per parameter, plus a fixed white-noise nugget equal to the
corresponding diagonal element of $\Gamma_{\mathrm{obs}}$, because each
training target is a single forward evaluation and not a replicate mean.
Hyperparameters are fitted by marginal likelihood with three restarts. The
predictive variance reported to the sampler is that of the latent mean map,
with the nugget excluded; including it would count $\Gamma_{\mathrm{obs}}$
twice in the likelihood and widen every posterior by about $\sqrt2$.

The training design is a stratified subsample of the calibration
evaluations from all twenty iterations, twenty-five particles per iteration,
$500$ points in total. This deviates from the published prescription, which
trains on the last iteration only ($100$ points), for the reason given in the
conjugate check below; the published choice is kept as the ablation configuration.

\subsection{Sampling}

The log posterior evaluated by the sampler is
\[
\log p(\theta \mid y) = \log \mathcal{N}\!\left(y \,\middle|\, m(\theta),\;
\Gamma_{\mathrm{obs}} + \mathrm{diag}\,v(\theta)\right) + \log \pi(\theta),
\]
where $m$ and $v$ are the emulator's predictive mean and latent variance and
$\pi$ is the truncated Gaussian prior, which is minus infinity outside the
box. It is sampled with the affine-invariant ensemble sampler of the
\texttt{emcee} package, with $32$ walkers, differential-evolution moves with
probability $0.8$ and snooker moves with probability $0.2$, initialized from
the final calibration ensemble with a jitter of $10^{-6}$ of the box, because
the emulator is informative only near its design and walkers started from the prior would sample regions where the emulator is unconstrained. Each chain runs
$1000$ burn-in and $4000$ production steps per walker, thinned by five, giving
$25{,}600$ kept draws per posterior. Acceptance fractions lie between $0.17$
and $0.37$ and integrated autocorrelation times between $11$ and $50$ steps
across all posteriors, so each chain holds several hundred effective samples
per walker.

\subsection{Flow posterior for the same observation}

The flow's posterior is computed for the same stored trajectory, cut to the window
and stride of the flow being used and passed through the conditioning it was
trained on, standardized with the training-split statistics saved by that run.
Of the three seeds of a configuration, the one with the lowest validation
loss is used, chosen without reference to the test split. Twenty thousand
draws are taken. Draws outside the prior's box are discarded before any
statistic is computed, because the prior gives them zero probability and the
sampler is already denied that region by its log prior; the discarded fraction
is recorded and is about one draw in twenty on Lorenz-96 and far fewer on
Lorenz-63. The same filter applied to the MCMC draws rejects none of them.

\subsection{Computation of reported statistics}

Every statistic is computed per parameter and per observation from the
marginal draws of one posterior, and the tables report averages over the eight
observations. Statistics are not averaged over parameters or computed jointly across them.

\begin{itemize}
\item \textbf{Posterior sd.} The standard deviation of the draws of that
  parameter. The tables give its mean over the eight observations, in the
  parameter's physical units.
\item \textbf{Bias over sd.} The posterior mean minus the truth, divided by
  the posterior sd. The tables give the mean of its absolute value over the
  eight observations. For a well calibrated Gaussian posterior this averages
  about $0.8$; larger values mean overconfidence.
\item \textbf{Coverage.} For each observation, whether the truth lies inside
  the central $90\%$ interval, that is between the $5$th and $95$th percentiles
  of the draws. The tables give the fraction of the eight observations for
  which it does, so coverage takes values in multiples of $0.125$ and a
  single observation moves it by that amount. With eight observations at a nominal $0.90$, six or fewer covered happens with probability $0.19$ and five or fewer with probability $0.04$, so only $0.63$ or below is evidence of overconfidence.
\item \textbf{Log score.} The logarithm of the marginal posterior density at
  the true value, estimated from the draws by a Gaussian kernel density with
  bandwidth $1.06\,\hat\sigma\,n^{-1/5}$, where $\hat\sigma$ is the standard
  deviation of the draws and $n$ their number, evaluated on at most $4000$
  evenly thinned draws. The tables give its mean over the eight observations.
  It is a density and not a probability, so it is not bounded by zero, its
  scale depends on the parameter's units, and it is comparable between methods
  for the same parameter but never across parameters. A difference of one
  between two methods means a density at the truth $2.7$ times higher.
\item \textbf{Composite parameters.} For $hc/b$ and $c\,b$ the composite is
  formed from every draw and the four statistics above are then computed from
  those transformed draws exactly as for a parameter. We do not form composites from posterior means.
\end{itemize}

\subsection{Verification on a conjugate problem}

Before any production run, the whole chain from calibration to sampling was
run on a linear forward map with a Gaussian prior, for which the posterior is
known in closed form. The ensemble Kalman sampler reproduces the exact
posterior mean and standard deviation to within ensemble noise, and the
likelihood evaluation matches an independent implementation to machine
precision. Three defects were found this way, each of which distorts the
posterior width without showing in the output on the real systems. First,
estimating $\Gamma_{\mathrm{obs}}$ at a stride different from the
observation's understates the mismatch variance. Second, adding the nugget to
the emulator's predictive variance widens every posterior by about $\sqrt2$.
Third, training the emulator on the last calibration iteration alone lets
posterior mass leak toward the prior, because the emulator's variance reaches
the size of $\Gamma_{\mathrm{obs}}$ within two to three posterior standard
deviations of the final ensemble and the likelihood flattens there; on the
conjugate problem this gave chain standard deviations three to four times the
exact value with the mean pulled toward the prior mean. Training on all
twenty iterations reproduces the exact posterior to $0.2$ standard deviations
in the mean and $2\%$ in the width, and is the production default.

\subsection{Cost}

Per observation, the calibration stage costs $2000$ forward runs, which is
about thirty minutes at the $100$ MTU Lorenz-96 window and six seconds on
Lorenz-63; fitting the emulators and running the chain adds roughly twenty
minutes. None of this is reusable for the next observation. The flow's
posterior for the same observation is one forward pass of about twenty
milliseconds after a training cost paid once for all observations.

\section{Additional CES posterior comparisons}
\label{sec:app_ces_extra}
The matched-statistics comparison evaluates differences between inference
procedures with observational inputs held fixed. The remaining tables and
figures report window dependence, composite-parameter constraints and
Lorenz-63 posterior quality for the eight held-out observations described in
Section~\ref{sec:ces}. These comparisons complement the aggregate accuracy
and coverage assessment in Appendix~\ref{sec:app_engines}.

\begin{table}[ht]
\color{black}
\centering\small
\begin{tabular}{@{}llcccccccc@{}}
\toprule
 & & \multicolumn{4}{c}{CES} & \multicolumn{4}{c}{flow, same statistics} \\
\cmidrule(lr){3-6}\cmidrule(lr){7-10}
window & quantity & $F$ & $h$ & $c$ & $b$ & $F$ & $h$ & $c$ & $b$ \\
\midrule
$2$ MTU   & posterior sd    & $0.73$ & $0.22$ & $1.62$ & $0.59$ & $0.60$ & $0.20$ & $1.48$ & $0.55$ \\
          & coverage $90\%$ & $0.88$ & $1.00$ & $0.88$ & $0.75$ & $1.00$ & $1.00$ & $0.88$ & $0.75$ \\
          & log score       & $-1.33$ & $0.32$ & $-1.90$ & $-1.02$ & $-0.80$ & $0.34$ & $-1.84$ & $-0.98$ \\
\addlinespace
$20$ MTU  & posterior sd    & $0.53$ & $0.17$ & $1.31$ & $0.49$ & $0.44$ & $0.17$ & $1.26$ & $0.51$ \\
          & coverage $90\%$ & $0.88$ & $0.88$ & $0.88$ & $0.88$ & $0.88$ & $0.88$ & $1.00$ & $0.88$ \\
          & log score       & $-0.77$ & $0.51$ & $-1.68$ & $-0.62$ & $-0.47$ & $0.09$ & $-1.54$ & $-0.99$ \\
\addlinespace
$100$ MTU & posterior sd    & $0.34$ & $0.13$ & $0.94$ & $0.35$ & $0.30$ & $0.12$ & $0.84$ & $0.35$ \\
          & coverage $90\%$ & $1.00$ & $0.88$ & $0.88$ & $0.75$ & $0.88$ & $0.75$ & $0.75$ & $0.75$ \\
          & log score       & $-0.09$ & $0.67$ & $-1.35$ & $-0.35$ & $-0.10$ & $0.04$ & $-1.60$ & $-0.90$ \\
\bottomrule
\end{tabular}
\caption{Lorenz-96, the matched control under the full
Calibrate--Emulate--Sample protocol, with eight held-out truths, each observed
through one stored trajectory. Both columns are conditioned on the same five
statistics of $(X,\Yb)$ over the same window, at $\Delta t = 1$ for $2$ MTU and
$\Delta t = 20$ otherwise. CES runs the ensemble Kalman sampler ($M = 100$,
twenty iterations) against the simulator, fits one Gaussian process per
statistic to the calibration evaluations, and samples the emulator by MCMC; the
flow is the summary-conditioned flow, best seed by validation loss. Posterior
sd and log score are averaged over the eight truths; coverage is the fraction
of truths inside the $90\%$ interval. The log score is the log posterior
density at the truth, comparable between methods for the same parameter only.}
\label{tab:ces_matched}
\end{table}

\begin{table}[ht]
\color{black}
\centering\small
\begin{tabular}{@{}llcccc@{}}
\toprule
method, conditioning & window & $F$ & $h$ & $c$ & $b$ \\
\midrule
CES, statistics of $(X,\Yb)$    & $2$ MTU   & $0.73$ & $0.22$ & $1.62$ & $0.59$ \\
CES, statistics of $(X,\Yb)$    & $20$ MTU  & $0.53$ & $0.17$ & $1.31$ & $0.49$ \\
CES, statistics of $(X,\Yb)$    & $100$ MTU & $0.34$ & $0.13$ & $0.94$ & $0.35$ \\
\addlinespace
flow, trajectory of $(X,\Yb)$ & $2$ MTU   & $0.27$ & $0.10$ & $0.77$ & $0.38$ \\
flow, trajectory of $(X,\Yb)$ & $20$ MTU  & $0.26$ & $0.10$ & $0.75$ & $0.34$ \\
flow, trajectory of $(X,\Yb)$ & $100$ MTU & $0.30$ & $0.09$ & $0.79$ & $0.27$ \\
\addlinespace
prior, for scale              &           & $1.32$ & $0.36$ & $1.91$ & $0.88$ \\
\bottomrule
\end{tabular}
\caption{Lorenz-96, posterior standard deviation as a function of window
length, mean over the eight held-out truths of Table~\ref{tab:ces_perobs}.
CES is conditioned on the five statistics of $(X,\Yb)$ over the stated window, at
$\Delta t = 1$ for $2$ MTU and $\Delta t = 20$ otherwise. The flow is
conditioned on the raw $(X,\Yb)$ trajectory over the stated window, at
$\Delta t = 1$ for $2$ MTU and $\Delta t = 20$ for $20$ and $100$ MTU, best seed
by validation loss. The last row is the standard deviation of each parameter
under the truncated prior.}
\label{tab:ces_curve}
\end{table}

\begin{table}[ht]
\color{black}
\centering\small
\begin{tabular}{@{}lcccccc@{}}
\toprule
 & \multicolumn{3}{c}{$hc/b$} & \multicolumn{3}{c}{$c\,b$} \\
\cmidrule(lr){2-4}\cmidrule(lr){5-7}
method, conditioning, window & sd & cov. & log score & sd & cov. & log score \\
\midrule
CES, statistics of $(X,\Yb)$, $2$ MTU    & $0.25$ & $0.88$ & $0.04$ & $16.2$ & $0.88$ & $-4.25$ \\
CES, statistics of $(X,\Yb)$, $20$ MTU   & $0.19$ & $1.00$ & $0.28$ & $12.7$ & $0.88$ & $-3.96$ \\
CES, statistics of $(X,\Yb)$, $100$ MTU  & $0.12$ & $1.00$ & $0.91$ & $8.8$  & $1.00$ & $-3.35$ \\
\addlinespace
flow, trajectory of $(X,\Yb)$, $2$ MTU   & $0.06$ & $0.88$ & $1.13$ & $5.5$ & $0.88$ & $-3.21$ \\
flow, trajectory of $(X,\Yb)$, $20$ MTU  & $0.07$ & $0.88$ & $1.44$ & $6.8$ & $1.00$ & $-3.17$ \\
flow, trajectory of $(X,\Yb)$, $100$ MTU & $0.06$ & $1.00$ & $1.67$ & $6.8$ & $1.00$ & $-2.95$ \\
\addlinespace
flow, trajectory of $X$ only, $2$ MTU    & $0.14$ & $0.88$ & $0.51$ & $18.8$ & $1.00$ & $-4.34$ \\
flow, statistics of $(X,\Yb)$, $100$ MTU   & $0.11$ & $0.88$ & $0.76$ & $7.9$  & $0.75$ & $-3.67$ \\
\addlinespace
prior, for scale                         & $0.44$ & & & $21$ & & \\
\bottomrule
\end{tabular}
\caption{Lorenz-96, the composite parameters per observation, eight held-out
truths. The composite of every posterior draw is taken, never the composite of
the means. Windows and sampling intervals as in Table~\ref{tab:ces_curve}. The
last two flow rows are the $X$-only trajectory flow at $\Delta t = 1$ and the
summary-conditioned flow of the matched control. The prior row is the standard
deviation of each composite under the truncated prior.}
\label{tab:ces_perobscomp}
\end{table}

\begin{table}[ht]
\color{black}
\centering\small
\begin{tabular}{@{}lccccccc@{}}
\toprule
 & \multicolumn{3}{c}{posterior sd} & cov. & \multicolumn{3}{c}{log score} \\
\cmidrule(lr){2-4}\cmidrule(lr){6-8}
conditioning & $\rho$ & $\sigma$ & $\beta$ & $\sigma$ & $\rho$ & $\sigma$ & $\beta$ \\
\midrule
CES, raw moments, $10$ MTU, $\Delta t = 1$            & $0.17$ & $0.88$ & $0.036$ & $1.00$ & $0.70$ & $-0.87$ & $2.25$ \\
\addlinespace
flow, same raw moments, $10$ MTU, $\Delta t = 1$      & $0.28$ & $0.65$ & $0.034$ & $0.88$ & $0.26$ & $-0.72$ & $2.26$ \\
flow, mean/sd/corr, $2$ MTU, $\Delta t = 1$           & $0.26$ & $0.27$ & $0.058$ & $1.00$ & $0.16$ & $0.25$ & $1.60$ \\
\addlinespace
flow, trajectory, $0.5$ MTU, $\Delta t = 1$           & $0.25$ & $0.26$ & $0.037$ & $1.00$ & $0.17$ & $0.15$ & $1.95$ \\
flow, trajectory, $2$ MTU, $\Delta t = 1$             & $0.21$ & $0.30$ & $0.038$ & $1.00$ & $0.55$ & $0.11$ & $2.11$ \\
flow, trajectory, $2$ MTU, $\Delta t = 20$            & $0.26$ & $0.34$ & $0.023$ & $0.88$ & $0.17$ & $-0.20$ & $2.75$ \\
flow, trajectory, $2$ MTU, $\Delta t = 50$            & $0.28$ & $0.36$ & $0.030$ & $0.75$ & $0.23$ & $-0.52$ & $2.36$ \\
flow, trajectory, $10$ MTU, $\Delta t = 20$           & $0.18$ & $0.38$ & $0.021$ & $0.63$ & $0.51$ & $-1.08$ & $2.70$ \\
flow, trajectory, $100$ MTU, $\Delta t = 20$          & $0.19$ & $0.28$ & $0.022$ & $1.00$ & $0.36$ & $0.12$ & $2.78$ \\
\addlinespace
flow, hybrid, $10$ MTU, $\Delta t = 20$               & $0.19$ & $0.18$ & $0.024$ & $1.00$ & $0.51$ & $0.50$ & $2.19$ \\
\addlinespace
prior, for scale                                      & $2.12$ & $1.25$ & $0.25$ & & & & \\
\bottomrule
\end{tabular}
\caption{Lorenz-63, per-observation comparison over eight held-out truths,
each observed through one stored trajectory. CES is conditioned on the nine raw
moments of the first $10$ MTU of the record at $\Delta t = 1$ ($T = 1000$).
Each flow is conditioned on the same record over the stated window and sampling
interval, as statistics, as the raw $(x,y,z)$ trajectory, or as the trajectory
together with its own statistics (hybrid); pooled encoder, best of three seeds
by validation loss. Coverage of the $90\%$ interval is shown for $\sigma$ only;
it is $1.00$ for $\rho$ in every row and at least $0.88$ for $\beta$ in every
row. Sampling intervals $\Delta t$ are expressed in stored steps of $0.01$ MTU,
and posterior standard deviations are averaged over eight observations.
The last row is the standard deviation of each parameter under the truncated prior.}
\label{tab:ces_perobs63}
\end{table}

\begin{figure}[ht]
\color{black}
\centering
\begin{subfigure}[t]{0.49\textwidth}\centering
\includegraphics[width=\textwidth]{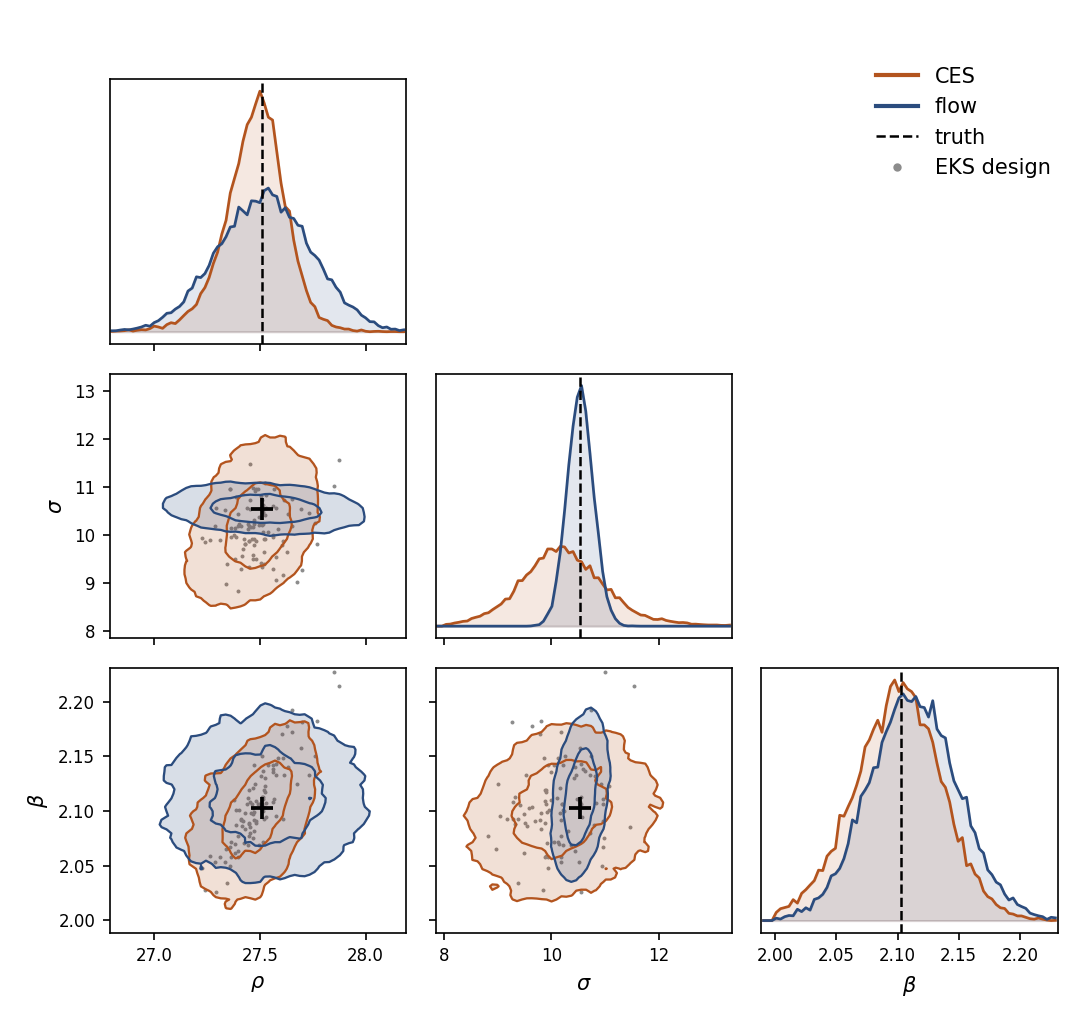}
\caption{Raw-trajectory flow, $2$ MTU, sampled every $0.01$ MTU.}
\end{subfigure}\hfill
\begin{subfigure}[t]{0.49\textwidth}\centering
\includegraphics[width=\textwidth]{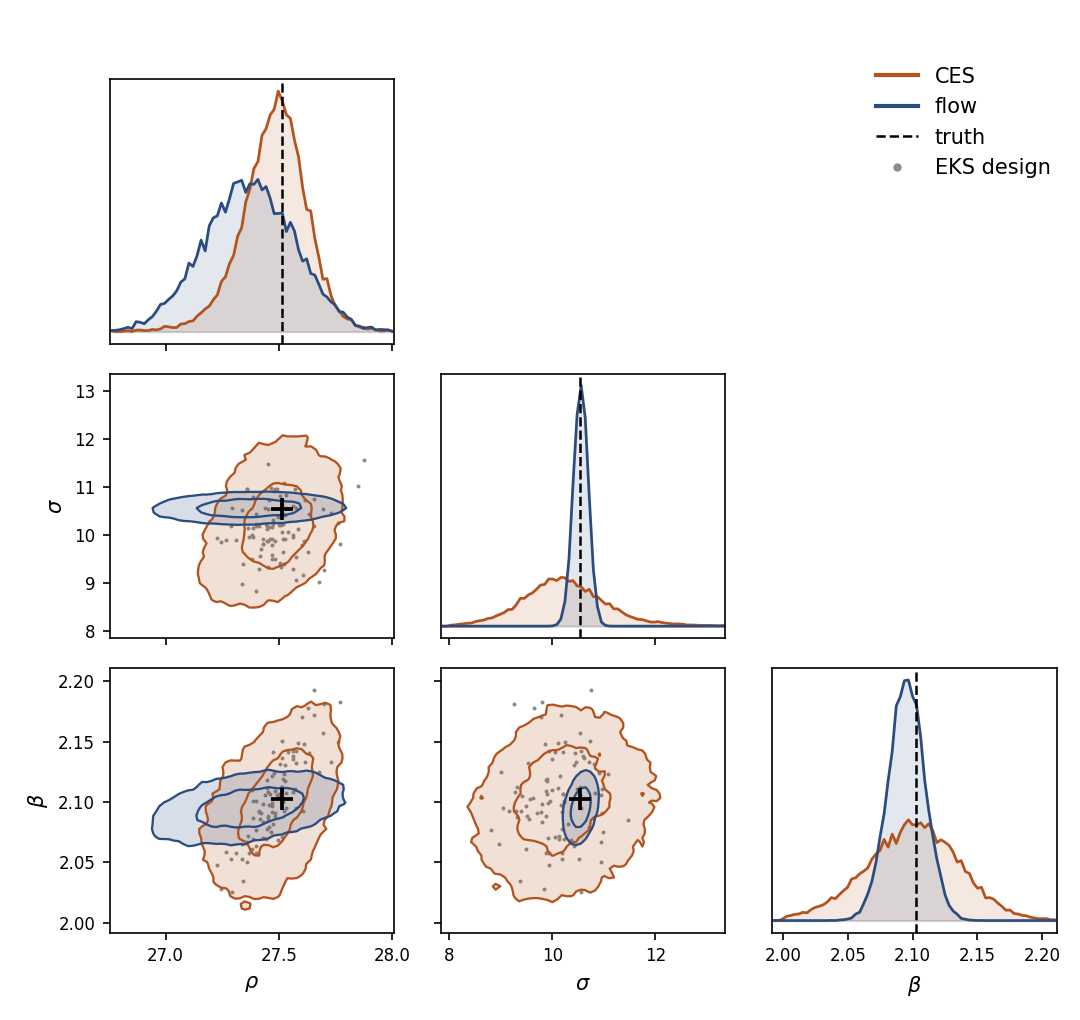}
\caption{Hybrid flow, $10$ MTU, sampled every $0.2$ MTU.}
\end{subfigure}
\caption{Lorenz-63, one held-out truth, $(\rho,\sigma,\beta) = (27.51, 10.53,
2.10)$. Layout as in Figure~\ref{fig:ces_perobs}. In both panels
Calibrate--Emulate--Sample (orange) is conditioned on the nine raw moments of
the first $10$ MTU sampled every $0.01$ MTU, with its last calibration ensemble in
grey. The flow (blue) is conditioned on the raw trajectory over the first $2$
MTU sampled every $0.01$ MTU (left), or on $10$ MTU sampled every $0.2$ MTU as
trajectory plus its own statistics (right, hybrid). Both use the same stored record. In both panels CES is
slightly sharper on $\rho$, the flow several times sharper on $\sigma$ and
centered on the truth, and both cover the truth on every parameter.}
\label{fig:ces_l63}
\end{figure}

\paragraph{Simulation accounting.}
Generating only the training records for a $2$ MTU flow requires about
$35{,}900$ realizations, each with $10$ MTU spinup, or $4.3\times10^5$ MTU
of integration. One $100$ MTU CES posterior requires about $2.2\times10^5$
MTU. Simulation costs are  comparable after roughly two CES
observations under this training-set-only accounting. It excludes validation
and test simulations, flow training, model selection and the longer records
generated for the full experimental program. Including spinup, replacing
$20$--$50$ MTU summary records with $2$ MTU trajectories reduces integration
per realization by a factor of $2.5$--$5$. All flow training runs in the full
program used about $140$ core-hours.

\section{Inference from identical summary statistics}
\label{sec:app_engines}

The shared-input comparison tests how much performance can differ between
inference procedures even without an encoder. At $1$ MTU on Lorenz-63,
$R^2_\sigma$ is $0.69$ for the summary flow and $0.18$ for the emulator,
with both giving near-nominal marginal coverage. Differences generally narrow
with longer records (Table~\ref{tab:ces_l63_engines}). Lorenz-96 likewise shows
parameter-dependent differences, with neither procedure consistently leading
at long windows (Table~\ref{tab:ces_l96_engines}). These results complement
the full CES matched-statistics control in Table~\ref{tab:ces_matched} and
show why gains from learned representations should be distinguished from
differences in approximate posterior inference.

\begin{table}[ht]
\color{black}
\centering\small
\begin{tabular}{@{}rrlrrrrrrr@{}}
\toprule
$T$ & MTU & engine & $R^2\,\rho$ & $R^2\,\sigma$ & $R^2\,\beta$ & cov $\rho$ & cov $\sigma$ & cov $\beta$ & s/obs \\
\midrule
$100$  & $1$  & GP emulator + MCMC & $0.866$ & $0.184$ & $0.290$ & $0.90$ & $0.86$ & $0.89$ & $21$ \\
       &      & flow               & $0.945$ & $0.686$ & $0.738$ & $0.92$ & $0.89$ & $0.89$ & $0.02$ \\
\addlinespace
$500$  & $5$  & GP emulator + MCMC & $0.987$ & $0.599$ & $0.914$ & $0.89$ & $0.91$ & $0.87$ & $22$ \\
       &      & flow               & $0.980$ & $0.773$ & $0.933$ & $0.94$ & $0.89$ & $0.91$ & $0.02$ \\
\addlinespace
$1000$ & $10$ & GP emulator + MCMC & $0.997$ & $0.752$ & $0.977$ & $0.91$ & $0.91$ & $0.87$ & $23$ \\
       &      & flow               & $0.978$ & $0.833$ & $0.977$ & $0.98$ & $0.90$ & $0.90$ & $0.02$ \\
\bottomrule
\end{tabular}
\caption{Lorenz-63, two inference engines given the same nine summary
statistics. $R^2$, coverage of the $90\%$ credible interval and
inference cost per observation on the same $153$ held-out parameter values
with ten stored trajectories each. Both engines are conditioned on the three
means and six raw second moments of the window at $\Delta t = 1$. The emulator
is one Gaussian process per statistic with a Mat\'ern-$5/2$ kernel, trained on
replicate-mean statistics at the $718$ training parameter values, followed by
affine-invariant ensemble MCMC with a Gaussian likelihood whose covariance is
the internal variability plus the emulator's predictive variance. The flow is
the summary-conditioned flow of Table~\ref{tab:l63_summary}, three-seed mean.
Cost per observation is inference only and excludes training. The flow's coverage
is computed over the same held-out rows from its in-support draws.}
\label{tab:ces_l63_engines}
\end{table}

\begin{table}[ht]
\color{black}
\centering\small
\begin{tabular}{@{}rlrrrrrr@{}}
\toprule
MTU & engine & $F$ & $h$ & $c$ & $b$ & $hc/b$ & s/obs \\
\midrule
$10$  & GP emulator + MCMC & $0.858$ & $0.804$ & $0.484$ & $0.735$ & $0.847$ & $1.91$ \\
      & flow               & $0.900$ & $0.841$ & $0.482$ & $0.782$ & $0.881$ & $0.021$ \\
\addlinespace
$50$  & GP emulator + MCMC & $0.929$ & $0.908$ & $0.699$ & $0.877$ & $0.922$ & $1.97$ \\
      & flow               & $0.952$ & $0.934$ & $0.719$ & $0.910$ & $0.946$ & $0.021$ \\
\addlinespace
$100$ & GP emulator + MCMC & $0.951$ & $0.920$ & $0.764$ & $0.897$ & $0.945$ & $2.03$ \\
      & flow               & $0.921$ & $0.942$ & $0.720$ & $0.923$ & $0.955$ & $0.021$ \\
\bottomrule
\end{tabular}
\caption{Lorenz-96, two inference engines given the same five summary
statistics. $R^2$ and inference cost per observation on the same $1530$
held-out rows. Both engines are conditioned on the site-and-window averages of
$X$, $\Yb$, $X^2$, $X\Yb$ and $\Yb^2$, the observation vector used by
Calibrate--Emulate--Sample, computed from every stored sample of the window.
The emulator and sampler are as in Table~\ref{tab:ces_l63_engines}; the flow
is the summary-conditioned flow. Coverage of the $90\%$ interval lies between
$0.89$ and $0.93$ for both engines at every window; the three-seed summary flows
of Table~\ref{tab:l96_fixedcost} at $10$, $50$ and $100$ MTU cover between
$0.87$ and $0.90$ on every parameter.}
\label{tab:ces_l96_engines}
\end{table}

\paragraph{Summary coordinates.}
Summary coordinates matter as well. Means, standard deviations and correlations encode the same information as means and raw second moments when variances are nonzero, yet approximate inference need not be invariant to that transformation. On an earlier Lorenz-63 dataset, switching to raw moments raised emulator $R^2_\beta$ from $0.64$ to $0.99$ at $10$ MTU (Table~\ref{tab:ces_coordinates}). Raw moments preserve the simple long-time balance $\langle xy\rangle = \beta\langle z\rangle$, while transformed coordinates express it through products of separately emulated quantities. This may explain some coordinate sensitivity. Flows are less sensitive here, though not invariant. Learning from trajectories reduces reliance on choosing both statistics and their coordinates in advance.

\section{Emulator baseline details}
\label{sec:app_emulator}
\begin{figure}[ht]
\color{black}
\centering
\includegraphics[width=0.66\textwidth]{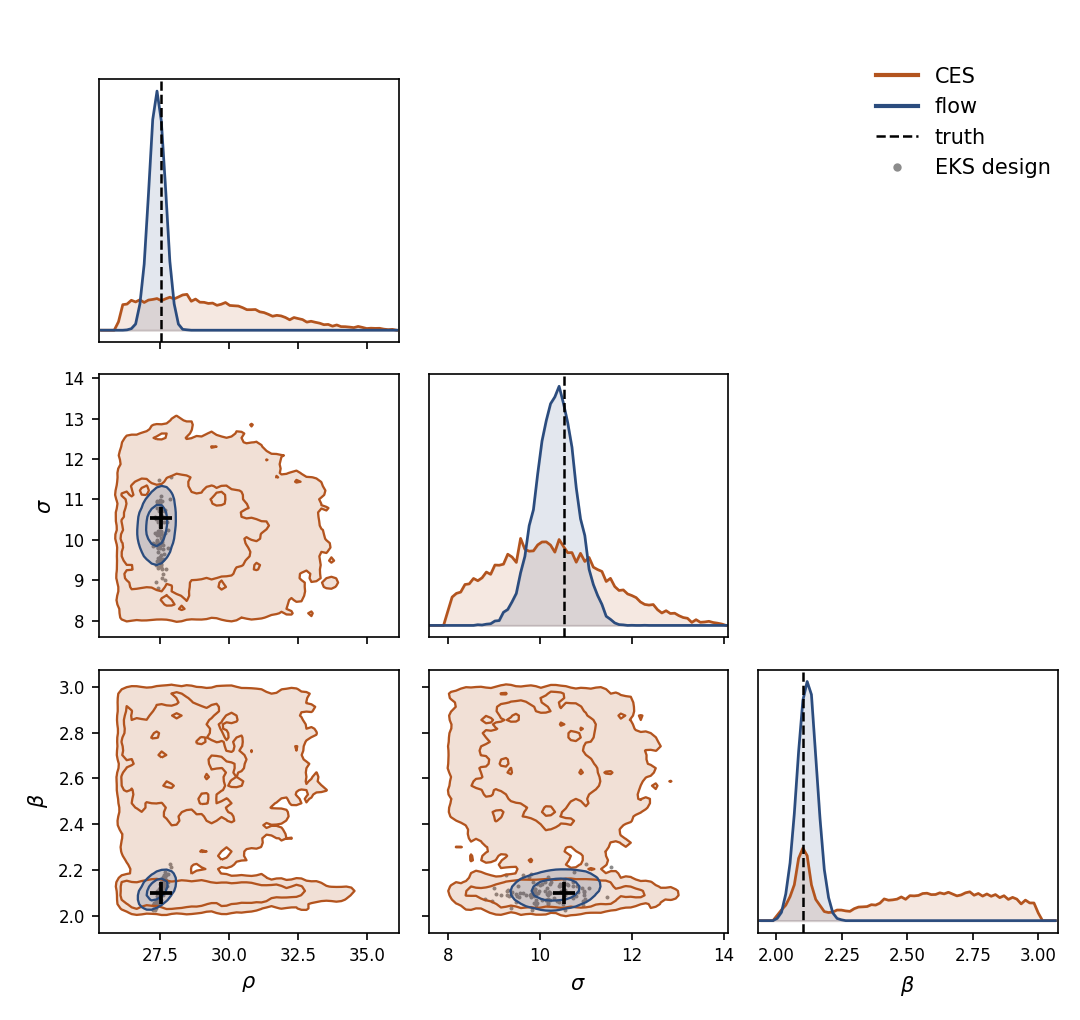}
\caption{The published emulator design on Lorenz-63 truth
$(\rho,\sigma,\beta) = (27.51, 10.53, 2.10)$ at $10$ MTU. Layout as in
Figure~\ref{fig:ces_perobs}. Calibrate--Emulate--Sample with its
emulator trained on the last calibration iteration only (orange) is shown with
the flow conditioned on the nine raw moments of the same window (blue). The calibration
ensemble (grey) sits on the truth, yet the orange posterior spreads across the
prior box in $\rho$ and, in $\beta$, splits into a peak at the truth and a
plateau over the rest of the prior. Away from the ensemble the emulator is
uninformed, the likelihood flattens, and posterior mass spreads toward the prior. 
Training the emulator on all twenty calibration iterations
(Table~\ref{tab:ces_ablation}) removes this effect.}
\label{fig:ces_ablation}
\end{figure}

\begin{figure}[ht]
\color{black}
\centering
\includegraphics[width=0.66\textwidth]{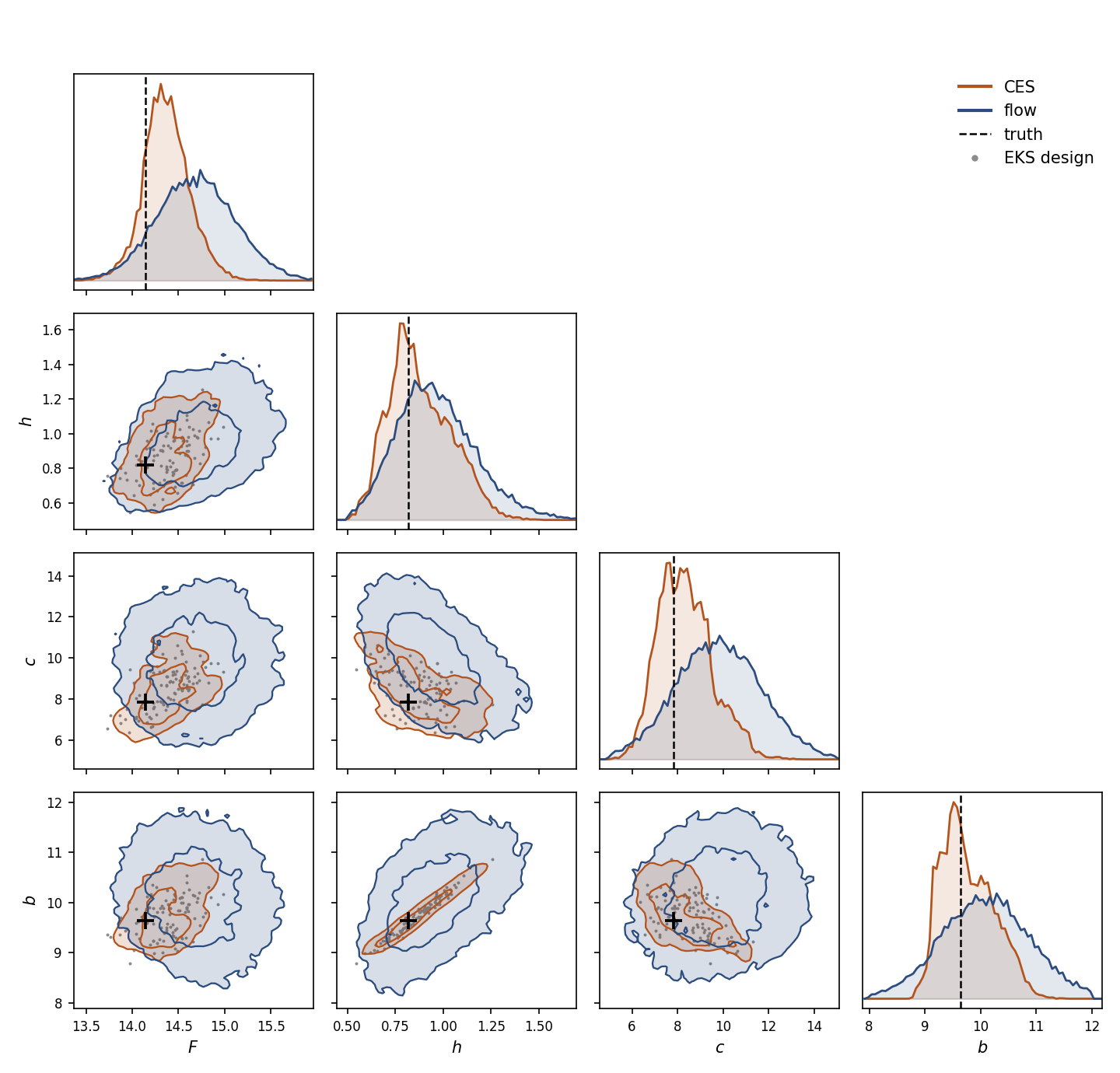}
\caption{Lorenz-96, the coupling degeneracy per observation, truth as in
Figure~\ref{fig:ces_perobs}. Layout as in Figure~\ref{fig:ces_perobs}. Calibrate--Emulate--Sample
conditioned on the five statistics of the full $100$ MTU record (orange) is
shown with the flow observing the slow variables $X$ only over the first $2$
MTU (blue). From $X$ alone the flow's posterior on $c$ and $b$ spans the prior
and follows a ridge in the $h$--$b$ plane, while its posterior on $F$ is as
sharp as that of CES.}
\label{fig:ces_xonly}
\end{figure}

Table~\ref{tab:ces_ablation} and Figure~\ref{fig:ces_ablation} measure the published emulator design against the one used in the main text. Table~\ref{tab:ces_coordinates} gives the dependence of the emulator on the coordinates in which the summary statistics are written. Table~\ref{tab:support} shows the effect of draws outside the prior support on a posterior mean. Figure~\ref{fig:ces_xonly} shows the coupling degeneracy per observation, with the flow observing $X$ alone against CES observing the statistics of $(X,\Yb)$ over $100$ MTU.

\begin{table}[ht]
\color{black}
\centering\small
\begin{tabular}{@{}llccc@{}}
\toprule
emulator design & quantity & $\rho$ & $\sigma$ & $\beta$ \\
\midrule
all iterations, $500$ points (this work) & posterior sd & $0.17$ & $0.88$ & $0.036$ \\
 & $|$bias$|$ / sd    & $0.63$ & $0.59$ & $0.46$ \\
 & coverage $90\%$    & $1.00$ & $1.00$ & $1.00$ \\
 & log score          & $0.70$ & $-0.87$ & $2.25$ \\
\addlinespace
last iteration, $100$ points (published prescription) & posterior sd & $1.98$ & $1.18$ & $0.19$ \\
 & $|$bias$|$ / sd    & $0.73$ & $0.66$ & $0.50$ \\
 & coverage $90\%$    & $1.00$ & $0.88$ & $1.00$ \\
 & log score          & $-1.34$ & $-1.43$ & $1.23$ \\
\addlinespace
prior, for scale & sd & $2.12$ & $1.25$ & $0.25$ \\
\bottomrule
\end{tabular}
\caption{Lorenz-63 at $10$ MTU, Calibrate--Emulate--Sample run with two
emulator designs on the same eight held-out observations. Both are conditioned
on the nine raw moments of the window at $\Delta t = 1$; the calibration stage,
the internal-variability estimate and the sampler are identical, and only the
set of calibration evaluations the emulator is trained on differs. The
all-iterations design is used throughout the main text; the last-iteration
design is the method's published prescription. The last-iteration posteriors are much wider, while the bias in units of posterior sd barely changes.}
\label{tab:ces_ablation}
\end{table}

\begin{table}[ht]
\color{black}
\centering\small
\begin{tabular}{@{}rrrr@{}}
\toprule
MTU & $\beta$ from $\langle xy\rangle/\langle z\rangle$ & flexible-regressor ceiling & emulator on mean/sd/correlation \\
\midrule
$5$  & $0.939$ & $0.954$ & $0.375$ \\
$10$ & $0.984$ & $0.976$ & $0.644$ \\
\bottomrule
\end{tabular}
\caption{Lorenz-63, the dependence of the emulator-and-sampler baseline on the
coordinates in which the summary statistics are written. $R^2$ for $\beta$
from the nine statistics of the window at $\Delta t = 1$. Column two evaluates
the exact relation $\beta = \langle xy\rangle/\langle z\rangle$, which follows
from averaging the $z$ equation, on the window's raw moments. Column three is a
gradient-boosted regressor from the nine statistics to $\beta$, an approximate
ceiling for any learner. Column four is the GP emulator plus MCMC conditioned
on the same nine numbers expressed as means, standard deviations and
correlations; re-expressed as raw moments, the same emulator reaches $0.989$ at
$10$ MTU. These numbers were obtained on an earlier Lorenz-63 dataset in a
non-canonical regime with $\rho$ near $56$, the only dataset on which the
correlation-coordinate emulator was run; the effect is a property of the
Gaussian likelihood and not of the regime.}
\label{tab:ces_coordinates}
\end{table}

\begin{table}[ht]
\color{black}
\centering\small
\begin{tabular}{@{}lrrrr@{}}
\toprule
estimator of the posterior center & $F$ & $h$ & $c$ & $b$ \\
\midrule
mean over all samples      & $0.916$ & $-3.9\times10^{20}$ & $-4.0\times10^{14}$ & $-3.2\times10^{9}$ \\
mean over in-support samples & $0.921$ & $0.942$ & $0.720$ & $0.923$ \\
median                     & $0.917$ & $0.943$ & $0.717$ & $0.920$ \\
\bottomrule
\end{tabular}
\caption{Lorenz-96 at $100$ MTU, $R^2$ of the summary-conditioned flow of
Table~\ref{tab:ces_l96_engines} under three estimators of the posterior center.
A normalizing flow pushes a Gaussian base through unbounded transformations, so
a small fraction of draws falls outside the truncated prior's support and a few
fall very far outside it. Restricting to in-support draws, or taking the
median, removes the contamination; the two agree to about $0.005$. All $R^2$
values in this paper use the in-support mean.}
\label{tab:support}
\end{table}

\section{Inequality check}
\label{sec:app_check}
A flow conditioned on the record can  do no worse, in information terms, than a flow conditioned on a summary of the same samples, on a subsample at a coarser interval, on a shorter prefix of the same window, or on a subset of its channels. Table~\ref{tab:check} applies this check to every trajectory-conditioned configuration for which such a flow exists. The benchmark on each parameter is the highest three-seed-mean score of any flow conditioned on a function of the record, including flows with a different encoder on the same record and, on Lorenz-96, a control flow on the site averages $\Yb$ alone. A flow fails on a parameter when it falls more than $0.05$ below a benchmark that itself exceeds $0.2$. A failure indicates a limitation of the trained estimator, since the record contains the benchmark's input.

Three patterns stand out. First, the raw-trajectory flows on Lorenz-63 fail in their three-seed means at every window because their seeds split, and even their best seeds fail at $0.05$ MTU. Second, on Lorenz-96 the flow on all sixty-four fast variables recovers less of $F$ ($0.59$) than a flow on their eight site averages alone ($0.71$), so the encoder misses a signal that averaging exposes. Third, the encoder sometimes reads a long sequence worse than a shorter or simpler view of the same record. The $X$ flow at $\Delta t = 1$ scores below the same window thinned to $\Delta t = 20$, the flows at $100$ MTU (hybrid on Lorenz-63, $(X,\Yb)$ on Lorenz-96) score below flows on their own first $10$ or $20$ MTU, and the pooled encoder on $\Yb$ alone scores below a final-state encoder on the same record, all in the three-seed mean. Every best seed outside the first two patterns passes, and the $(X,\Yb)$ flows at $2$ and $20$ MTU pass in the mean as well.

\begin{table}[ht]
\color{black}
\centering\scriptsize
\setlength{\tabcolsep}{3pt}
\begin{tabular}{@{}llllll@{}}
\toprule
input & window, $\Delta t$ & three-seed mean & best seed & benchmark & fails (mean / best) \\
\midrule
\multicolumn{6}{@{}l}{\textit{Lorenz-63}, parameters $\rho/\sigma/\beta$} \\
$(x,y,z)$, trajectory & $0.05$, $1$ & $0.68/0.35/0.30$ & $0.81/0.48/0.72$ & $0.93/0.90/0.91$ & $\rho,\sigma,\beta$ / $\rho,\sigma,\beta$ \\
$(x,y,z)$, trajectory & $0.5$, $1$ & $0.86/0.56/0.61$ & $0.98/0.97/0.97$ & $0.93/0.94/0.91$ & $\rho,\sigma,\beta$ / none \\
$(x,y,z)$, trajectory & $2$, $1$ & $0.96/0.30/0.32$ & $1.00/0.97/0.99$ & $0.99/0.94/0.98$ & $\sigma,\beta$ / none \\
$(x,y,z)$, trajectory & $10$, $20$ & $0.99/0.28/0.78$ & $0.99/0.92/1.00$ & $0.99/0.96/0.99$ & $\sigma,\beta$ / none \\
$(x,y,z)$, trajectory & $100$, $20$ & $0.98/0.31/0.45$ & $0.99/0.96/1.00$ & $0.99/0.96/0.99$ & $\sigma,\beta$ / none \\
$(x,y,z)$, hybrid & $10$, $20$ & $0.99/0.96/0.99$ & $0.99/0.99/0.99$ & $0.99/0.49/0.94$ & none / none \\
$(x,y,z)$, hybrid & $100$, $20$ & $0.95/0.88/0.98$ & $0.97/0.91/0.99$ & $0.99/0.96/0.99$ & $\sigma$ / none \\
\addlinespace
\multicolumn{6}{@{}l}{\textit{Lorenz-96}, parameters $F/h/c/b$} \\
$X$, trajectory & $2$, $1$ & $0.89/0.61/0.12/0.00$ & $0.95/0.66/0.12/0.04$ & $0.95/0.68/0.13/0.06$ & $F,h$ / none \\
$X$, trajectory & $2$, $5$ & $0.90/0.66/0.12/0.06$ & $0.96/0.68/0.14/0.06$ & $0.95/0.68/0.13/0.06$ & none / none \\
$X$, trajectory & $20$, $20$ & $0.96/0.70/0.12/0.06$ & $0.97/0.71/0.12/0.08$ & $0.96/0.71/0.13/0.06$ & none / none \\
$X$, trajectory & $100$, $20$ & $0.96/0.71/0.13/0.08$ & $0.96/0.71/0.13/0.08$ & $0.96/0.71/0.13/0.06$ & none / none \\
$X$, hybrid & $100$, $20$ & $0.95/0.69/0.13/0.06$ & $0.97/0.71/0.12/0.08$ & $0.96/0.71/0.13/0.08$ & none / none \\
$Y$, trajectory & $2$, $1$ & $0.59/0.93/0.88/0.79$ & $0.60/0.93/0.90/0.79$ & $0.71/0.94/0.88/0.78$ & $F$ / $F$ \\
$Y$, trajectory & $2$, $5$ & $0.52/0.92/0.78/0.74$ & $0.60/0.92/0.80/0.76$ & $0.38/0.84/0.23/0.45$ & none / none \\
$Y$, trajectory & $20$, $20$ & $0.83/0.97/0.80/0.90$ & $0.84/0.97/0.83/0.90$ & $0.71/0.96/0.36/0.90$ & none / none \\
$\Yb$, trajectory & $2$, $1$ & $0.55/0.80/0.59/0.67$ & $0.67/0.88/0.73/0.74$ & $0.71/0.87/0.70/0.71$ & $F,h,c$ / none \\
$(X,\Yb)$, trajectory & $2$, $1$ & $0.94/0.90/0.72/0.76$ & $0.96/0.92/0.77/0.77$ & $0.96/0.92/0.70/0.74$ & none / none \\
$(X,\Yb)$, trajectory & $2$, $5$ & $0.96/0.92/0.70/0.74$ & $0.96/0.92/0.68/0.74$ & $0.93/0.82/0.52/0.48$ & none / none \\
$(X,\Yb)$, trajectory & $2$, $20$ & $0.93/0.82/0.52/0.48$ & $0.94/0.82/0.52/0.49$ & $0.95/0.68/0.13/0.06$ & none / none \\
$(X,\Yb)$, trajectory & $20$, $20$ & $0.96/0.90/0.73/0.80$ & $0.96/0.92/0.79/0.84$ & $0.96/0.83/0.56/0.71$ & none / none \\
$(X,\Yb)$, trajectory & $100$, $20$ & $0.94/0.86/0.57/0.84$ & $0.97/0.94/0.80/0.91$ & $0.96/0.90/0.73/0.82$ & $c$ / none \\
\bottomrule
\end{tabular}
\caption{The inequality check. For each trajectory-conditioned configuration,
$R^2$ of the three-seed mean and of the seed with the lowest validation loss,
against the benchmark, which on each parameter is the highest three-seed-mean
score of any flow conditioned on a function of the same record (a summary of
the same samples, a coarser subsample, a shorter prefix, a subset of channels,
or a different encoder on the same record). Windows are in MTU and $\Delta t$
in steps of $0.01$ MTU. A parameter is listed as failing when the flow falls
more than $0.05$ below a benchmark above $0.2$. Values are rounded to two
decimals from the three-decimal values of the other tables, so $1.00$ means at
least $0.995$.}
\label{tab:check}
\end{table}
\color{black}
\end{document}